\documentclass[a4paper]{aa}

\usepackage{graphicx,float}
\usepackage{tabularx}
\usepackage{latexsym,amsmath,amssymb}
\usepackage{natbib}
\bibpunct{(}{)}{;}{a}{}{,}

\def \arvind    {A.~N. Parmar}

\def \laurence  {L. Boirin}

\def \mariano {M. M{\'e}ndez}
\def \maria {M. D{\'i}az Trigo}
\def \jelle {J.~S. Kaastra}

\def \sron {The SRON, the National Institute for Space Research,
  Sorbonnelaan 2, 3584 CA Utrecht, The Netherlands}

\def \strasbourg {Observatoire Astronomique de Strasbourg, 11 rue de
l'Universit\'e, F-67000 Strasbourg, France}

\def \estec {Astrophysics Missions Division, Research and Scientific Support
                Department of ESA, ESTEC, Postbus 299, NL-2200 AG
                Noordwijk, The Netherlands}

\def\countsec{\hbox{counts s$^{-1}$}}

\def\degmark{^\circ}
\def \rsun {\ifmmode$R$_{\odot}\else R$_{\odot}$}

\def \flux {$F_{\rm X}$}
\def \lum {$L_{\rm X}$}

\def \hcm {\hbox {\ifmmode $ atoms cm$^{-2}\else atoms cm$^{-2}$\fi}}

\def\approxgt{\mathrel{\hbox{\rlap{\lower.55ex \hbox {$\sim$}}
        \kern-.3em \raise.4ex \hbox{$>$}}}}
\def\approxlt{\mathrel{\hbox{\rlap{\lower.55ex \hbox {$\sim$}}
        \kern-.3em \raise.4ex \hbox{$<$}}}}

\newcommand {\msun} {{{\rm M$_{\odot}$}}}

\def\arcsec{\hbox{$^{\prime\prime}$}}
\newcommand {\ergs} {erg~s$^{-1}$}
\newcommand {\ergcms} {erg cm$^{-2}$ s$^{-1}$}
\newcommand {\chisq} {$\chi ^{2}$}
\newcommand {\rchisq} {$\chi_{\nu} ^{2}$}

\newcommand {\phind} {$\Gamma$} 
\newcommand {\ecut} {$E_{\rm c}$}

\newcommand {\egau} {$E_{\rm gau}$}

\newcommand {\ktbb} {$kT_{\rm bb}$}

\newcommand {\sig} {$\sigma$}
\newcommand {\ew} {$EW$}
\newcommand {\ews} {$EW$s}

\def \logn {$\log(N)$}

\def \nhabs {$N{\rm _H^{abs}}$}
\def \nhxabs {$N{\rm _H^{xabs}}$}

\newcommand {\ttnh} {$\times~$10$^{22}$~atoms~cm$^{-2}$}

\newcommand {\fetfour} {\ion{Fe}{xxiv}}
\newcommand {\fetfive} {\ion{Fe}{xxv}}
\newcommand {\fetsix} {\ion{Fe}{xxvi}}
\newcommand {\nitseven} {\ion{Ni}{xxvii}}
\newcommand {\ssixteen} {\ion{S}{xvi}}
\newcommand {\sifourteen} {\ion{Si}{xiv}}
\newcommand {\sithirteen} {\ion{Si}{xiii}}

\newcommand {\catwenty} {\ion{Ca}{xx}}
\newcommand {\fetone} {\ion{Fe}{xxi}}
\newcommand {\fettwo} {\ion{Fe}{xxii}}
\newcommand {\fetthree} {\ion{Fe}{xxiii}}
\newcommand {\canineteen} {\ion{Ca}{xix}}
\newcommand {\arseventeen} {\ion{Ar}{xvii}}
\newcommand {\sfifteen} {\ion{S}{xv}}
\newcommand {\oeight} {\ion{O}{viii}}
\newcommand {\neten} {\ion{Ne}{x}}
\newcommand {\mgtwelve} {\ion{Mg}{xii}}
\newcommand {\niteight} {\ion{Ni}{xxviii}}

\newcommand {\dptd} {$\rm ^{2}P_{3/2}$}
\newcommand {\dpud} {$\rm ^{2}P_{1/2}$}
\newcommand {\upu} {$\rm ^{1}P_{1}$}
\newcommand {\tpu} {$\rm ^{3}P_{1}$}

\def \xiunit {\hbox{erg cm s$^{-1}$}}
\def \logxi {$\log(\xi)$}

\def \xabs {{\tt xabs}}
\newcommand {\sigmav} {$\sigma_{\rm v}$}
\newcommand {\kms} {km~s$^{-1}$}

\def \kpl {$k_{\rm pl}$}
\def \kbb {$k_{\rm bb}$}
\def \kgau {$k_{\rm gau}$}

\def \nineteen {XB\,1916$-$053}
\def \mxb {MXB\,1658$-$298}
\def \bigdip {X\,1624$-$490}
\def \twelve {X\,1254$-$690}
\def \grs {GRS\,1915+105}
\def \gro {GRO\,J1655$-$40}
\def \gx {GX\,13+1}
\def \cir {Cir\,X$-$1}

\def \thirteen {4U\,1323$-$62}
\def \seventeen {H\,1743$-$322}

\def \src {\thirteen}

\begin{document}

\title{A highly-ionized absorber in the X-ray binary \src: a~new
  explanation for the dipping phenomenon}

\author{\laurence\inst{1, 2} \and \mariano\inst{1} \and \maria\inst{3} \and \arvind\inst{3} \and \jelle\inst{1}}

\offprints{L. Boirin, \email{boirin@astro.u-strasbg.fr}}

\institute{\sron \and \strasbourg \and \estec}

\date{Received 3 September 2004 / Accepted 15 February 2005}

\authorrunning{L. Boirin et al.}

\titlerunning{XMM-Newton observation of \src}

\abstract{ We report the detection of narrow \fetfive\ and \fetsix\
  X-ray absorption lines at $6.68 \pm 0.04$~keV and $6.97 \pm
  0.05$~keV in the persistent emission of the dipping low-mass X-ray
  binary \src\ during a 2003 January XMM-Newton observation. These
  features are superposed on a broad emission feature centered on
  $6.6\,^{+0.1}_{-0.2}$~keV. During dipping intervals the equivalent
  width of the \fetfive\ feature increases while that of the \fetsix\
  feature decreases, consistent with the presence of less strongly
  ionized material in the line-of-sight.  As observed previously, the
  changes in the 1.0--10~keV spectrum during dips are inconsistent
  with a simple increase in absorption by cool material.  However, the
  changes in both the absorption features {\it and} the continuum can
  be modeled by variations in the properties of an ionized absorber.
  No partial covering of any component of the spectrum, and hence no
  extended corona, are required.  From persistent to deep dipping the
  photo-ionization parameter, $\xi$, expressed in \xiunit, decreases
  from \logxi\ of $3.9 \pm 0.1$ to \logxi\ of $3.13 \pm 0.07$, while
  the equivalent hydrogen column density of the ionized absorber
  increases from ($3.8 \pm 0.4$)~\ttnh\ to ($37 \pm 2$)~\ttnh.  Since
  highly-ionized absorption features are seen from many other dip
  sources, this mechanism may also explain the overall changes in
  X-ray spectrum observed during dipping intervals from these systems.
   }

\maketitle

\section{Introduction}
\label{sec:intro}

\src\ is a faint (3--10 mCrab) low-mass X-ray binary (LMXB) which
exhibits X-ray bursts and 2.94~hour periodic intensity dips. The
source was first detected by {\it Uhuru} and {\it Ariel V}
\citep{forman78apj,warwick81mnras} and the dips and bursts were
discovered using EXOSAT \citep{1323:vdk85ssr,1323:parmar89apj}.  From
the burst properties, the distance of \src\ was constrained to be
between 10--20~kpc \citep{1323:parmar89apj}. During the dips, which
typically last for 30\% of the orbital cycle, the 1--8 keV intensity
varies irregularly with a minimum of $\sim$50\% of the average value
outside of the dips. These dips are believed to be due to obscuration
in the thickened, azimuthally structured, outer regions of an
accretion disk
\citep{1916:white82apjl}.  The presence of periodic dips indicates
that the source is viewed relatively close to edge-on, at an
inclination angle, $i$, of 60--80$\degmark$
\citep{frank87aa}, where $i$ is defined as the angle between the
line-of-sight and the rotation axis of the accretion disk.

The BeppoSAX 1.0--150~keV spectrum of the \src\ persistent emission
can be modeled by a cutoff power-law with a photon index, \phind, of
$1.48 \pm 0.01$ and a folding energy, \ecut, of $44.1 \,
^{+5.1} _{-4.4}$~keV together with a blackbody with a temperature of
$1.77 \pm 0.25$~keV which contributes $\sim$15\% of the 2--10~keV flux
\citep{1323:balucinska99aa}. From observations with the RXTE
Proportional Counter Array, \citet{1323:barnard01aa} report the
presence of a $6.43 \pm 0.21$~keV Fe emission feature with an
equivalent width, $EW$, of $110 \pm 55$~eV.  Quasi-periodic
oscillations at $\sim$1~Hz were discovered in the persistent emission,
dips, and bursts by \citet{1323:jonker99apjl}.  Radial intensity
profiles from the imaging Medium-Energy Concentrator Spectrometer on
BeppoSAX indicate the presence of a substantial dust halo around the
source \citep{1323:barnard01aa}.  Such an intense halo is not
unexpected since \src\ is located close to the galactic plane ($b =
0.5\degmark$) and its emission is absorbed by interstellar material in
the line-of-sight equivalent to $4 \times
10^{22}$~H~atoms~cm$^{-2}$. This high absorption probably also
explains why the IR counterpart is undetected at optical wavelengths
\citep{1323:smale95aj}.

The 1--10~keV spectra of most of the dip sources, including \src,
become harder during dipping.  However, these changes are inconsistent
with a simple increase in photo-electric absorption by cool material,
as an excess of low-energy photons is usually present. Two approaches
have been used to model this spectral evolution. Initially, in the
``absorbed plus unabsorbed'' approach
\citep[e.g.,][]{0748:parmar86apj} the persistent (non-dipping)
spectral shape was used to model spectra from dipping intervals. It
was included with, and without, additional absorption. The spectral
evolution during dipping was accounted for by a large increase in the
column density of the absorbed component, and a decrease of the
normalization of the unabsorbed component. The latter decrease was
attributed to electron scattering in the absorber.  More recently, in
the ``progressive covering'', or ``complex continuum'' approach
\citep[e.g.,][]{1916:church97apj}, the X-ray emission is assumed to
originate from a point-like blackbody, or disk-blackbody component,
together with a power-law component from an extended corona.  This
approach models the spectral changes during dipping intervals by the
partial and progressive covering of the extended component by an
opaque neutral absorber.  The absorption of the point-like component
is allowed to vary independently from that of the extended component.
This approach has been successfully applied to a number of dipping
LMXBs including \src\
\citep[][]{1323:balucinska99aa,1323:barnard01aa}.

The improved sensitivity and spectral resolution of {\it Chandra} and
XMM-Newton is allowing narrow absorption features from highly ionized
Fe and other metals to be observed from a growing number of X-ray
binaries.  In particular, \fetfive\ (He-like) or \fetsix\ (H-like)
1s-2p resonant absorption lines near 7~keV were reported from the
micro-quasars \gro\ \citep{1655:ueda98apj,1655:yamaoka01pasj}, \grs\ 
\citep{1915:kotani00apj,1915:lee02apj} and \seventeen\ 
\citep{1743:miller04apj}, and from the neutron star systems \cir\ 
\citep{cirx1:brandt96mnras,cirx1:brandt00apjl,cirx1:schulz02apj}, \gx\ 
\citep{gx13:ueda01apjl,gx13:sidoli02aa,gx13:ueda04apj}, \mxb\ 
\citep{1658:sidoli01aa}, \bigdip\ \citep{1624:parmar02aa}, \twelve\ 
\citep{1254:boirin03aa} and \nineteen\ \citep{1916:boirin04aa}.  The
lines from \cir, \gx\ and \seventeen\ are blue-shifted, indicating
that the highly ionized plasma is an outflow in these sources. The
systems that exhibit the \fetfive\ and \fetsix\ features are mainly
dipping sources \citep[see Table~5 of ][]{1916:boirin04aa}.  The lack
of any orbital phase dependence of the features (except during dips)
suggests that the absorbing plasma is located in a thin cylindrical
geometry around the compact object.  Such highly ionized plasma is
probably a common feature of accreting binaries, but is preferentially
detected in the dipping sources, presumably due to being viewed at
inclination angles within $\sim$$30\degmark$ of the orbital plane.
 
Here, we report the detection of narrow \fetfive\ and \fetsix\ 1s-2p
X-ray absorption lines from the dipping LMXB \src. These features are
independently reported by \citet{1323:church05mnras}. We show that the
lines are associated with an ionized absorber which has different
properties during persistent and dipping intervals.  We demonstrate
that these differences can account for the broad-band spectral changes
observed between persistent and dipping intervals.

\section{Observation and data analysis}
\label{sec:reduction}

\subsection{Data reduction}

The XMM-Newton Observatory \citep{jansen01aa} includes three
1500~cm$^2$ X-ray telescopes each with an European Photon Imaging
Camera (EPIC) at the focus.  Two of the EPIC imaging spectrometers use
MOS CCDs \citep{turner01aa} and one uses PN CCDs \citep{struder01aa}.
Reflection Grating Spectrometers \citep[RGS,][]{denherder01aa} are
located behind two of the telescopes.  \src\ was observed by
XMM-Newton for 50~ks on 2003 January 29 between 09:05 and 22:58~UTC.
The thin optical blocking filter was used with the EPIC cameras.  The
EPIC PN and MOS1 cameras were operated in timing mode.  The EPIC MOS2
camera was operated in full window mode.  As the MOS2 full window mode
data is strongly affected by pile-up and MOS1 timing mode data is
currently not well calibrated, we concentrate on the analysis of PN
data.  We note that the PN CCD is more sensitive to the presence of
lines than the MOS CCDs with an effective area a factor $\sim$5 higher
at 7~keV and a better energy resolution.  RGS data from both gratings
in first and second order were also used.  All data products were
obtained from the XMM-Newton public archive and reduced using the
Science Analysis Software (SAS) version 5.4.1.  No intervals of
enhanced solar activity were present.

In PN timing mode, only one PN CCD chip (corresponding to a field of
view of 13\farcm6$\times$4\farcm4) is used and the data from that chip
are collapsed into a one-dimensional row (4\farcm4) to be read out at
high speed.  This allows a time resolution of 30~$\mu$s, and photon
pile-up occurs only for count rates $>$1500~\countsec. Only single and
double events (patterns 0 to 4) were selected. Source events were
extracted from a 53\arcsec\ wide column centered on the source
position (RAWX 30 to 43).  Background events were obtained from a
column of the same width, but centered 115\arcsec\ from \src\ (RAWX 2
to 15).  The latest response matrix file for the PN timing mode
provided by the XMM-Newton calibration team (epn\_ti40\_sdY9.rsp,
released in 2003 January), was used.  Ancillary response files were
generated using the SAS task {\tt arfgen}.  EPIC PN spectra were
rebinned to oversample the full-width at half-maximum ($FWHM$) of the
energy resolution by a factor 3, and to have a minimum of 25 counts
per bin to allow the use of the $\chi^2$ statistic.

%
%
\begin{figure}[!th]
\centerline{\includegraphics[angle=270,width=0.5\textwidth]{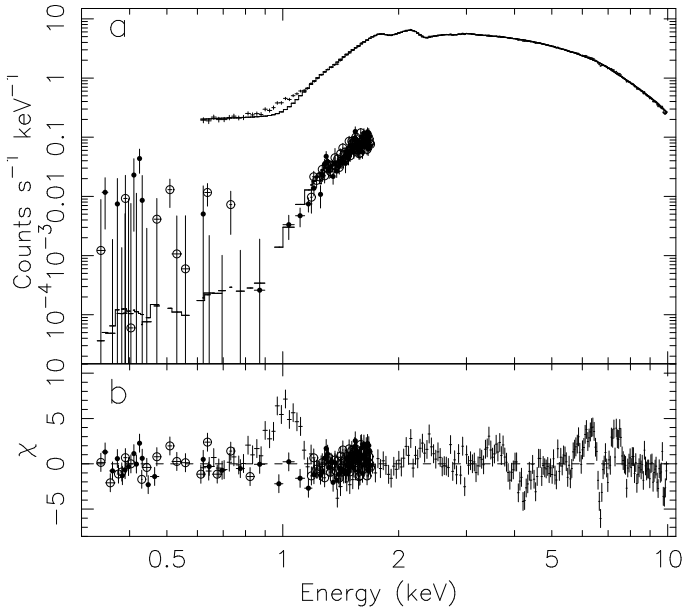}}
\caption[]{{\bfseries a)} RGS and PN spectra of the entire \src\ observation.
First order RGS1 and RGS2 spectra are plotted using empty and filled
circles, respectively. The second order spectra are consistent with
the first order, but are not shown for clarity. The stepped line shows
the best-fit absorbed blackbody plus power-law model.  {\bfseries b)}
Residuals from the best-fit model in units of standard deviations.
There is good agreement between the PN and RGS in the overlapping
energy range, except near 1~keV.}
\label{fig:rgspndiscrepant}
\end{figure}

The SAS task {\tt rgsproc} was used to produce calibrated RGS event
lists, spectra, and response matrices.  The RGS spectra do not show
evidence for any narrow spectral features.  Therefore, they were
mainly used to check for consistency with the PN spectra and to better
constrain the low-energy part of the spectrum. They were rebinned to
have a minimum of 25 counts per bin.

\subsection{Spectral analysis}
\label{sec:spectralanalysis}

In order to check for consistency between the different instruments,
we first examined the spectrum of the entire observation. Since this
includes persistent, dipping and bursting intervals, it is of limited
scientific interest.  We combined the four 0.3--1.7~keV RGS spectra
(RGS1 and RGS2, orders 1 and 2) and the 0.6--10~keV PN spectrum.  We
fit them with a model consisting of a blackbody and a power-law, both
modified by photo-electric absorption from neutral material.  A
constant factor, fixed to 1 for the PN spectrum, but allowed to vary
for each RGS spectrum, was included multiplicatively in order to
account for cross-calibration uncertainties. The residuals from the
best-fit model are shown in Fig.~\ref{fig:rgspndiscrepant}. The fit is
unacceptable (reduced \chisq, \rchisq, of 2.7 for 431 degrees of
freedom, d.o.f.), due mainly to the presence of strong absorption
lines near 7~keV superposed on a broad emission feature and a
significant excess in the PN spectrum near 1~keV, which is not
detected in the RGS. Since this feature may have an instrumental
origin, we concentrate on the spectral structure near 7~keV in this
paper.

Therefore, we use EPIC PN spectra only in the energy range
1.7--10~keV. We include RGS spectra (RGS1 and RGS2, orders 1 and 2) to
better constrain the low-energy spectral shapes of the average of the
persistent, shallow and deep dipping intervals (Sect.
\ref{sec:averaged}).  We restrict the RGS energy range to
1.0--1.7~keV because very few source events are detected below 1~keV.
When the data are further divided (Sect.~\ref{sec:timeresolved}), the
RGS count rate of the individual segments becomes too low for the
spectra to be useful and only PN data are used.  In either case, since
there is no evidence for any narrow features in the RGS spectra, only
the PN spectra are plotted to improve clarity.

Spectral analysis was performed using XSPEC \citep{arnaud96conf}
version 11.2, and SPEX package \citep{kaastra96conf} version
2.00.09. The photo-electric cross sections of
\citet{morrison83apj} are used throughout to account for absorption by
neutral gas with solar abundances \citep[][{\tt abs} model within
SPEX, and {\tt wabs} model in XSPEC]{anders89gca}.  Spectral
uncertainties are given using $\Delta$\chisq of 2.71, corresponding to
90\% confidence for one interesting parameter, and to 95\% confidence
for upper limits. The
\chisq\ values obtained from SPEX are calculated using estimated
errors on the model rather than errors on the data (see the SPEX
user's manual). All $EW$s are quoted with positive values both for
absorption and emission features.

%
%
\begin{figure*}[!ht]
\centerline{\hspace{-1cm}\includegraphics[angle=90,width=1.15\textwidth]{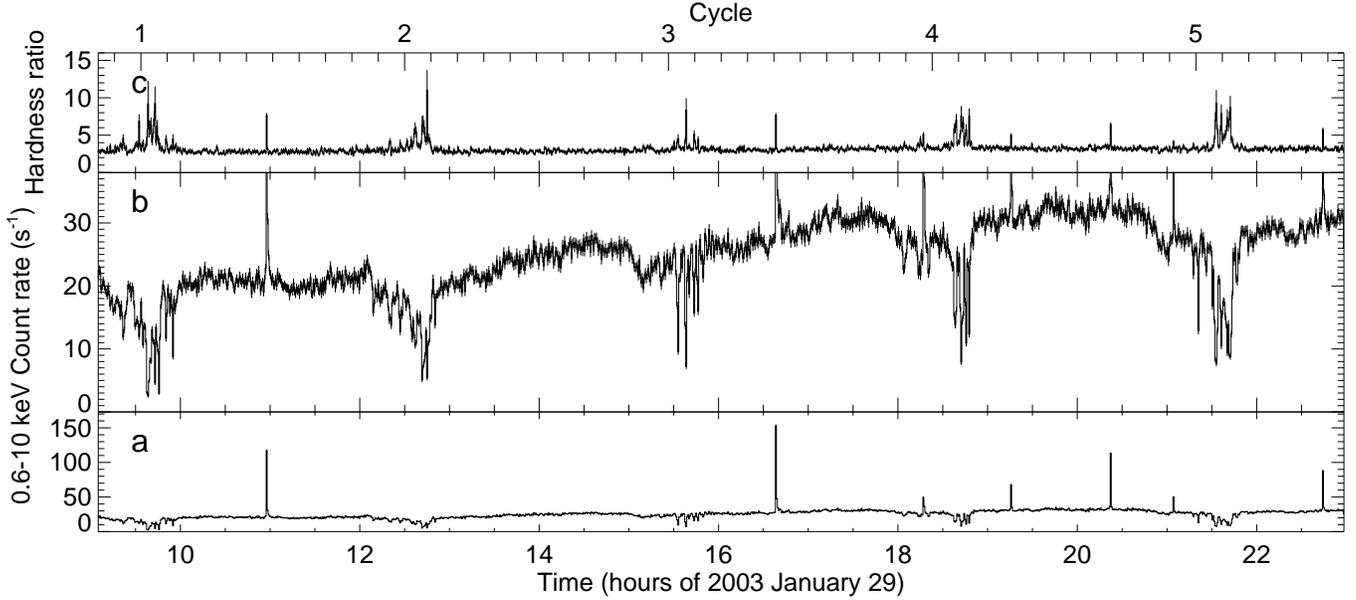}}
\caption{
  {\bfseries a)} 0.6--10~keV EPIC PN lightcurve of \src\ showing 7
  X-ray bursts and 5 dips. {\bfseries b)} Only the low intensity part
  of the lightcurve is shown (the bursts are truncated). {\bfseries
  c)} Hardness ratio (counts in the 2.5--10~keV band divided by those
  between 0.6--2.5~keV). The cycle number is indicated on the top
  axis; integer values correspond to phase 0 or estimated dip
  mid-times. The reference time was chosen as 9.57~hr on 2003 January
  29 and the period is 2.938~h. The binning is 40~s in each panel.}
\label{fig:lc}
\end{figure*}

%
%
\begin{figure*}[!th]
\centerline{\hspace{-0.3cm}\includegraphics[angle=0,width=0.38\textwidth]{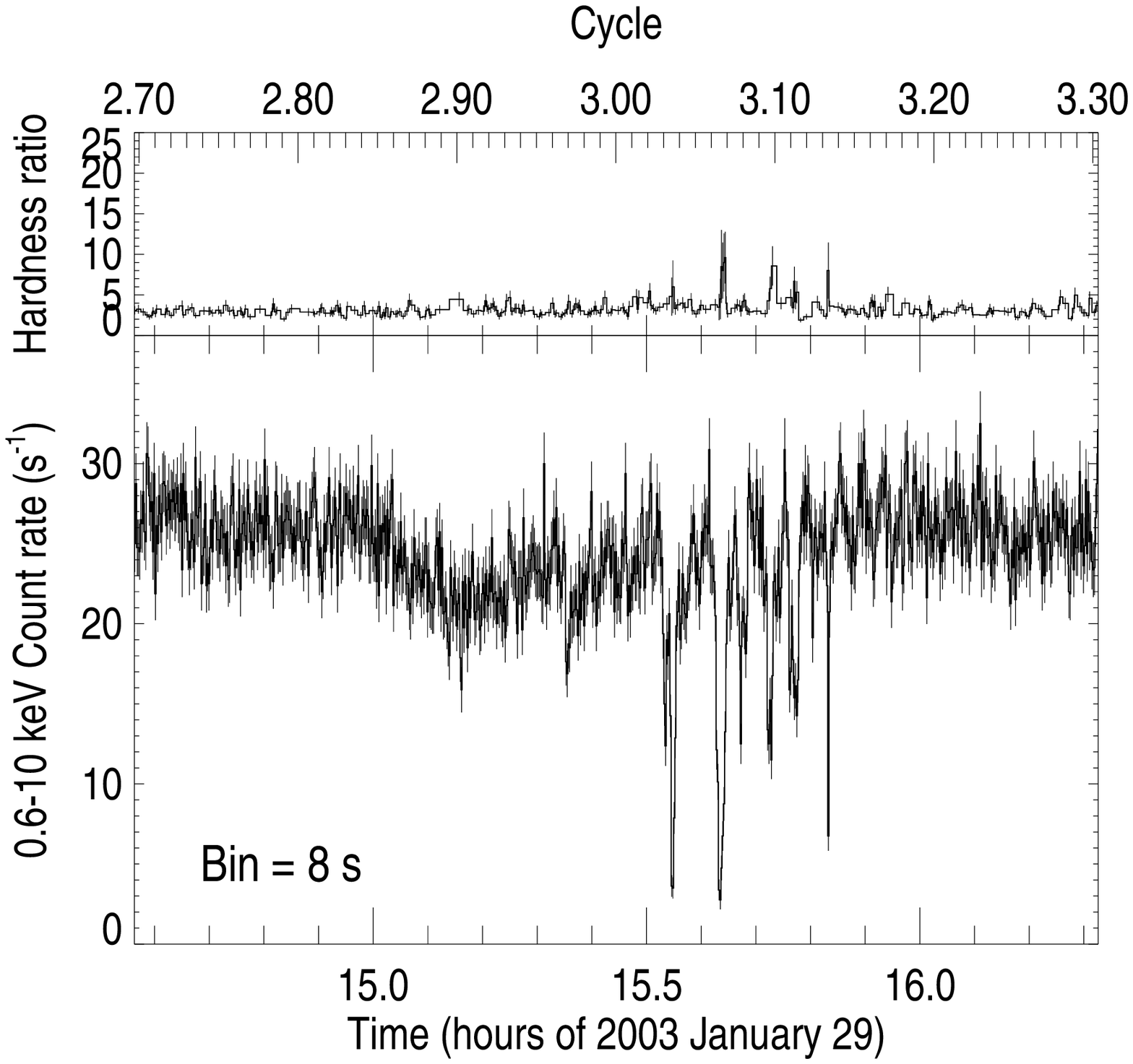}\hspace{-0.8cm}\includegraphics[angle=0,width=0.38\textwidth]{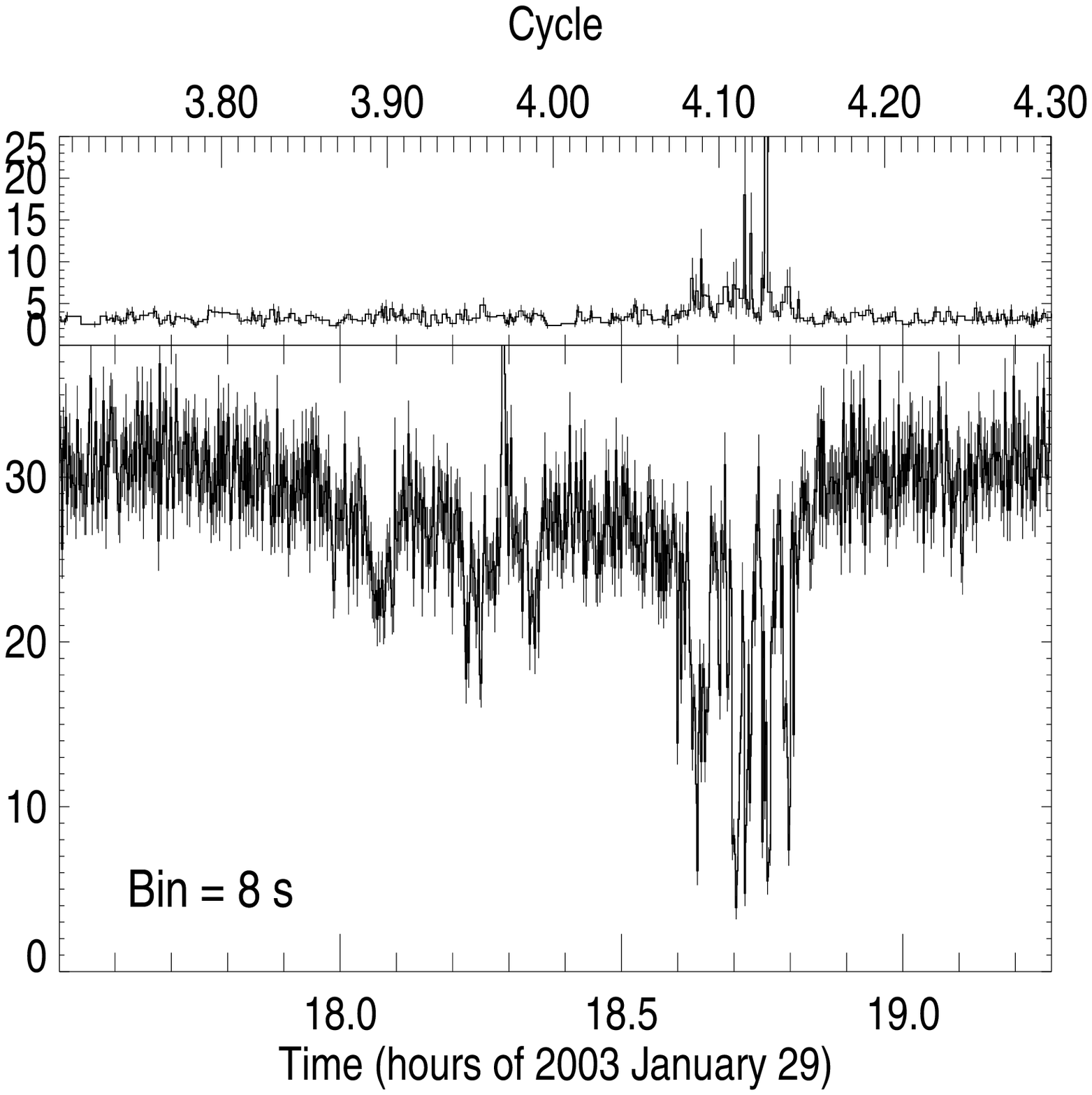}\hspace{-0.1cm}\includegraphics[angle=0,width=0.135\textwidth]{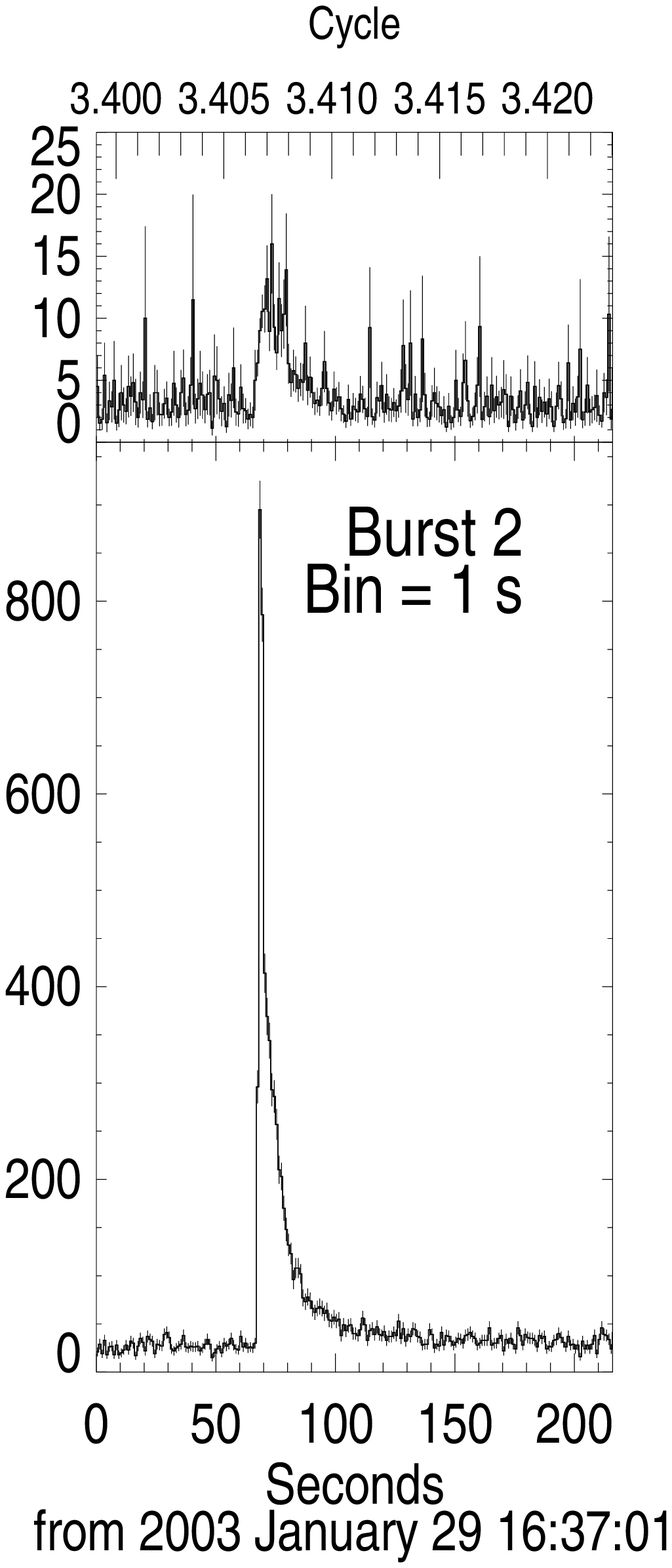}\hspace{0.1cm}\includegraphics[angle=0,width=0.135\textwidth]{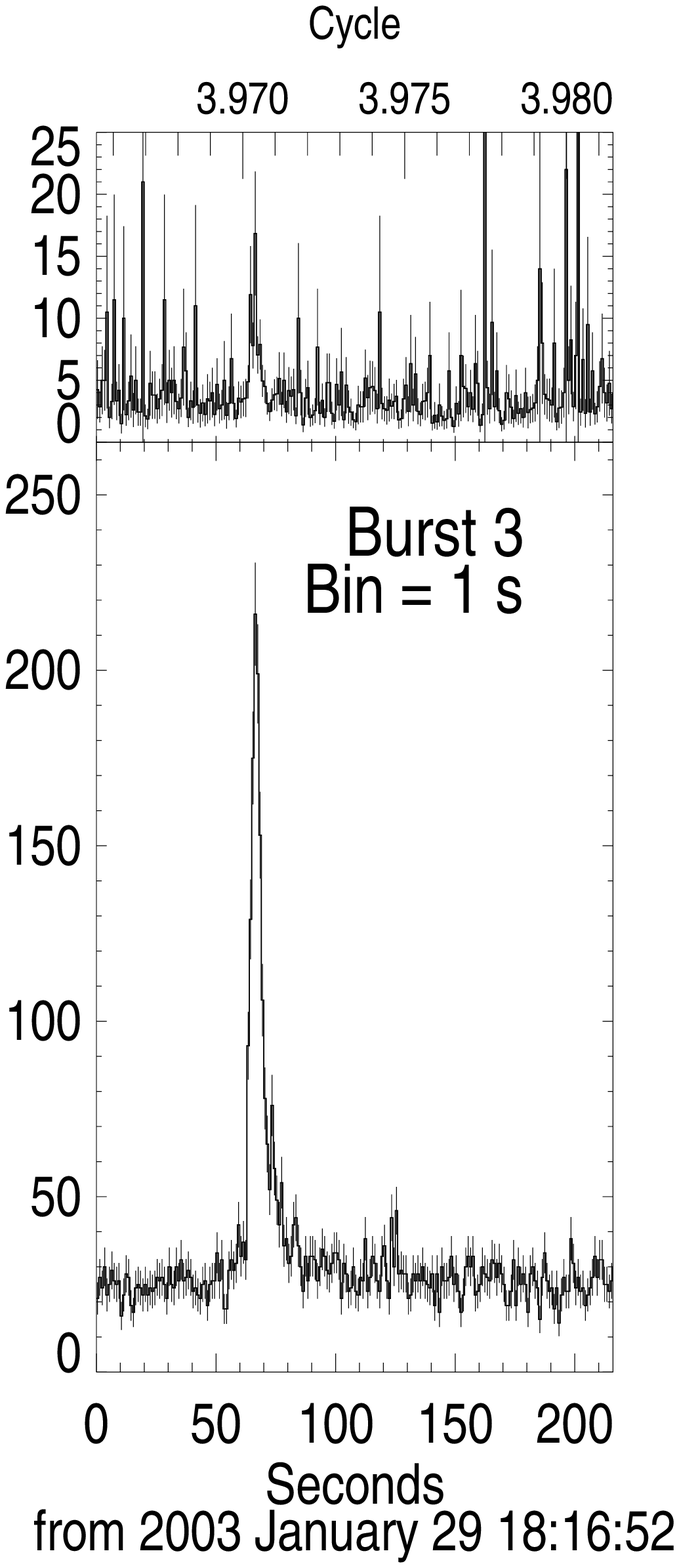}}
\caption[]{
  0.6--10~keV EPIC PN lightcurves of \src\ showing two dips with a
  binning of 8~s and two bursts with a binning of 1~s. Rapid intensity
  variability is visible during the dips. Burst 3 (right) occurs
  during dipping activity and is much fainter than burst 2 which
  occurs during a persistent interval.  The upper panels show the
  hardness ratio and the top axis the cycle number, both defined as in
  Fig.~\ref{fig:lc}.}
\label{fig:zoomlc}
\end{figure*}

\subsection{Narrow absorption features}
\label{sec:narrow}

The narrow absorption features evident near 7~keV were accounted for
by including a photo-ionized absorber in the spectral model. We use
the \xabs\ model of SPEX, which treats, in a simplified manner, the
absorption by a thin slab composed of different ions, located between
the ionizing source and the observer.  The \xabs\ model assumes that
the angle subtended by the slab as seen from the ionizing source is
small. Therefore, emission from the slab and scattering by the slab of
the ionizing source into the line-of-sight are neglected, and only
absorption and scattering out of the line-of-sight by the slab are
considered.  The processes taken into account are the continuum and
the line absorption by the ions and (from SPEX version 2.00.09)
scattering out of the line-of-sight by the free electrons in the slab.
The transmission, $T$, of the slab is calculated as $T =
\exp[-\tau_{\rm c} -\tau_{\rm l} -\tau_{\rm e}]$, where $\tau_{\rm c}$
and $\tau_{\rm l}$ are the total continuum and line optical depth,
respectively, and $\tau_{\rm e}$ is the electron scattering optical
depth.  The relativistic energy-dependent Klein-Nishina correction is
applied to compute $\tau_{\rm e}$, but the classical Thomson
approximation is valid below 10~keV where we will examine the
spectra. Therefore, $\tau_{\rm e} \sim \sigma_{\rm T}N_{\rm e}$, with
$\sigma_{\rm T}$ the Thomson cross-section and $N_{\rm e}$ the
electron column density, and this scattering term is energy
independent. On the contrary, $\tau_{\rm c}$ and $\tau_{\rm l}$ are
energy dependent.  Most continuum opacities are taken from
\citet{verner95aa}, while line opacities and wavelengths for most ions
are taken from \citet{verner96ad} (see the details and additional
references in the SPEX user's manual).  $\tau_{\rm l}$ is a function
of $v$ and \sigmav\ which are free parameters of \xabs. $v$ is the
average systematic velocity shift of the absorber, in \kms. Negative
and positive values of $v$ correspond to blue and red shifts,
respectively.  \sigmav\ is the turbulent velocity broadening of the
absorber in \kms, defined as $\sigma_{\rm total}^{2} = \sigma_{\rm v
}^{2} + \sigma_{\rm thermal}^{2}$, where $\sigma_{\rm total}$ is the
total width of a line and $\sigma_{\rm thermal}$ the thermal
contribution.  $\tau_{\rm l}$ is also a function of the optical depth
of each individual line, $\tau_{\rm i}$, and thus of the column
density of each ion.  However, in \xabs, the relative column densities
of the ions are coupled through a photo-ionization model, so that only
two free parameters are introduced: \nhxabs\ and $\xi$.  \nhxabs\ is
the equivalent hydrogen column density of the ionized absorber in
units of atoms~cm$^{-2}$.  $\xi$ is the ionization parameter of the
absorber defined as $\xi = L /n_{\rm e} \, r^{2}$, where $L$ is the
luminosity of the ionizing source, $n{\rm _e}$ the electron density of
the plasma and $r$ the distance between the slab and the ionizing
source. $\xi$ is expressed in units of \xiunit, but we will omit the
units when quoting \logxi\ values in this paper.  Using codes such as
XSTAR \citep{kallman01apjs} or CLOUDY \citep{ferland93araa}, and
assuming a broad-band ionizing continuum from infra-red to hard
X-rays, the ionic column densities of a photo-ionized slab can be
pre-calculated for different values of $\xi$.  During the fitting
process, SPEX reads in the grid of pre-calculated ionic column
densities and finds the best set and consequently the best-fit values
for \nhxabs\ and $\xi$. $N_{\rm e}$ and $\sigma_{\rm thermal}$ are
linked self-consistently to these two parameters, respectively, via
the photo-ionization model.
  
In this paper, we use a grid of ionic column densities pre-calculated
using CLOUDY and assuming that the ionizing continuum may be
represented by a cutoff power-law with \ecut\ of 44~keV, as measured
in the BeppoSAX spectrum of \src\ \citep{1323:balucinska99aa}, and
with \phind\ of 1.96. This value of \phind\ was obtained by fitting
simultaneously the EPIC PN and RGS persistent spectra of \src\ by an
absorbed cutoff power-law with \ecut\ fixed to 44~keV (although better
fits are obtained with multi-component modeling, see
e.g. Sect.~\ref{sec:averaged}).

The approach followed using \xabs\ can be considered as a first
attempt to include physics when modeling the absorption features.
Although the assumed geometry is a very simple one, a strong advantage
of \xabs\ is that all relevant ions are automatically taken into
account and their relative column densities are coupled in a physical
way.  Since ions having small cross-sections can contribute
significantly to the absorption when combined, and since absorption
features can be blended, \xabs\ represents a strong improvement over
the approach consisting in accounting for each apparent absorption
feature individually in a spectrum. Nevertheless, in order to give a
more objective description of the absorption features, we also fit the
most prominent ones with Gaussian line profiles using XSPEC.

The effects of assuming a different ionizing continuum on our results
 and the importance of line re-emission into the line-of-sight are
 estimated in Appendix~A.

%
%
\begin{table}[!t]
\begin{center}
\caption[]{Cycles (as defined in Fig.~\ref{fig:lc}) corresponding to
  persistent, shallow dipping and deep dipping intervals.}
\begin{tabular}{ccc}
\hline
\hline
\noalign {\smallskip}
Persistent 	& \multicolumn{2}{c}{Dipping}\\
	 	& Shallow 	& Deep \\
\noalign {\smallskip}\hline
	        & 0.83--1.01 & 1.01--1.08  \\
1.20--1.86 	& 1.86--2.02 & 2.02--2.10 \\
2.12--2.85 	& 2.85--3.02 & 3.02--3.12  \\
3.20--3.80 	& 3.80--4.08 & 4.08--4.15  \\
4.20--4.80 	& 4.80--5.06 & 5.06--5.14  \\
5.20--5.60 	& &\\
\noalign {\smallskip}
\hline
\end{tabular}
\label{tab:timeselection}
\end{center}
\end{table}

%
%
\begin{figure}[!ht]
\centerline{\hspace{-0.5cm}\includegraphics[angle=90,width=0.55\textwidth]{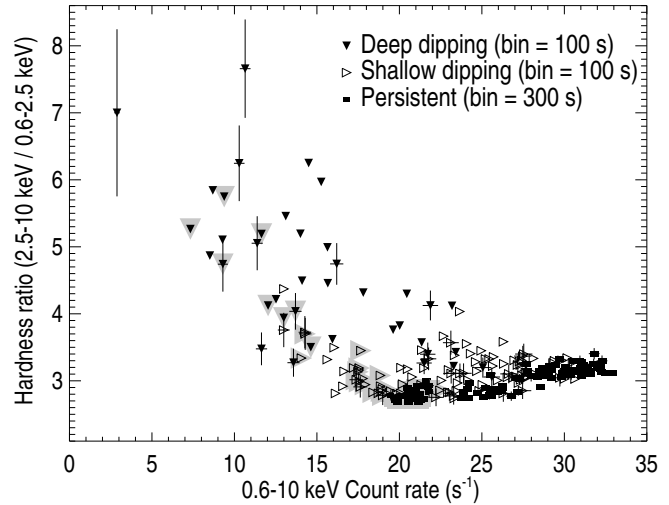}}
\caption{Hardness-intensity diagram. Filled rectangles, 
  empty and filled triangles indicate the persistent, shallow and deep
  dipping intervals, respectively. We show in grey the track followed
  by \src\ from persistent emission to deep dipping over one single
  cycle (from cycle 1.20 to 2.10). The secular evolution of the
  persistent emission throughout the entire observation is visible as
  the track formed by the filled rectangles: the persistent hardness
  ratio is correlated to the count rate.  Error bars are given on a
  few points. }
\label{fig:hrint}
\end{figure}

\section{Results}

\subsection{Lightcurve and hardness-intensity diagram}

The EPIC PN 0.6--10~keV lightcurve is shown in Fig.~\ref{fig:lc}
(panels a and b) with a binning of 40~s. Panel c shows the hardness
ratio (counts in the 2.5-10~keV energy range divided by those between
0.6--2.5~keV) also with a binning of 40~s. Times are not
barycenter-corrected. The cycle number is indicated on the top axis.
This was determined from the reference time of 9.57~hr on 2003 January
29 (corresponding to an XMM-Newton time of 160209470.4~s) visually
estimated as the dip center time, and using a period of 2.938~hr
\citep{1323:balucinska99aa}.  Integer values of the cycle number
correspond to phase 0, or estimated dip mid-times.  This convention
differs from the one adopted in X-ray binaries showing both periodic
dips and total eclipses, where phase 0 is usually chosen as the mid
eclipse time. Fig.~\ref{fig:hrint} shows the hardness ratio as a
function of the 0.6--10~keV count rate.

%
%
\begin{figure*}[ht!]
\centerline{\includegraphics[width=0.37\textwidth]{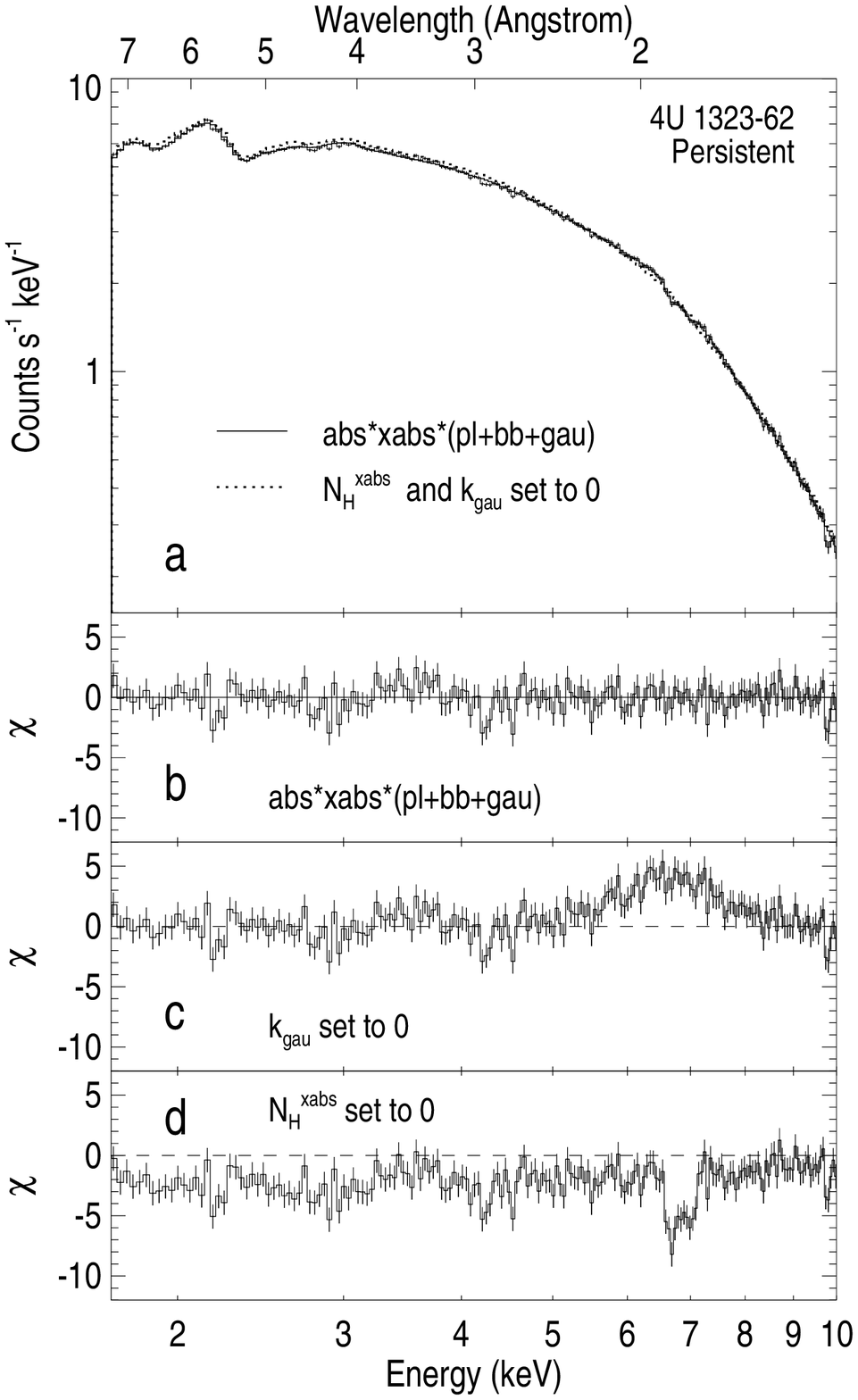}
\hspace{-1.1cm}
\includegraphics[width=0.37\textwidth]{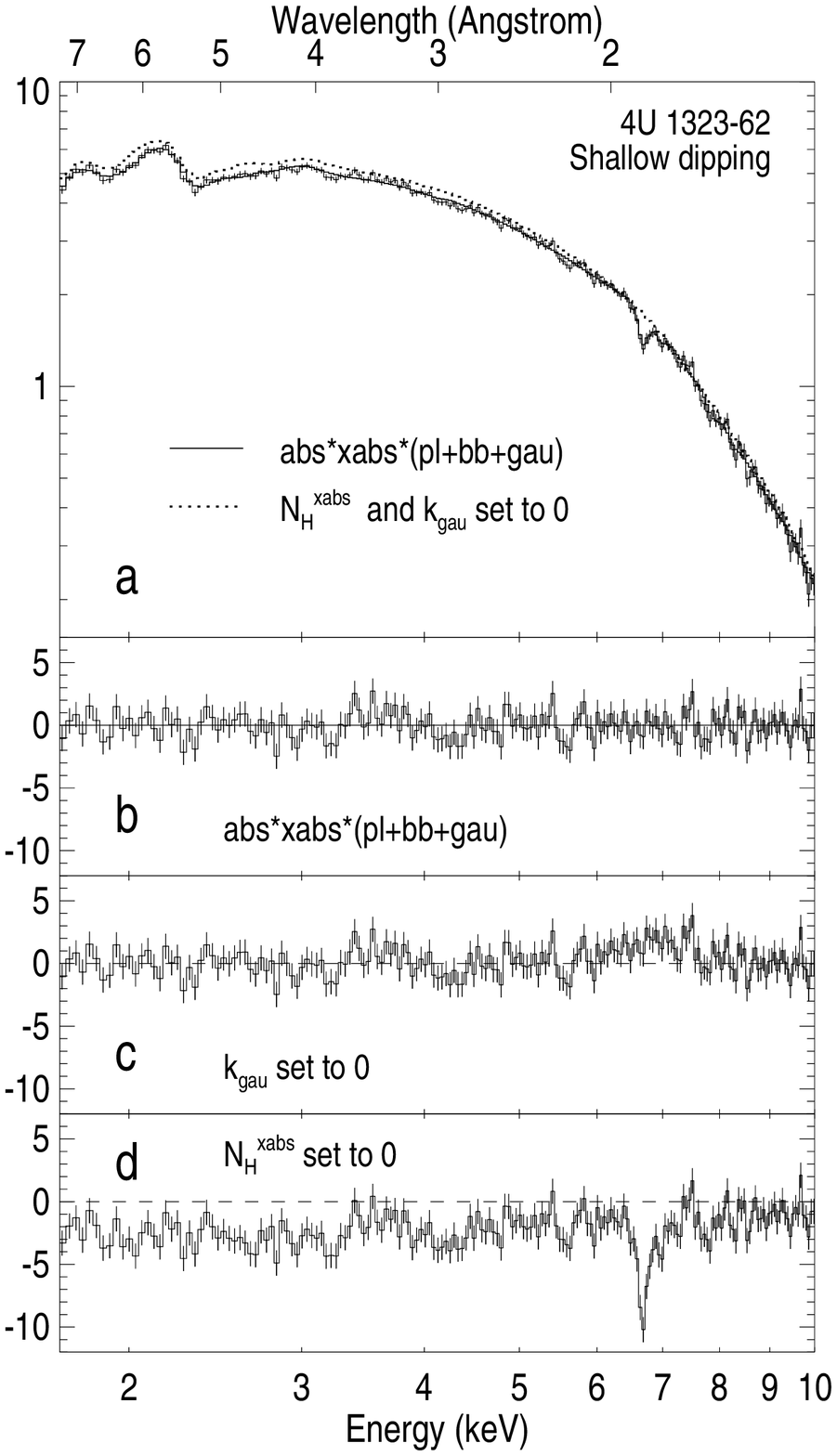}
\hspace{-1.1cm}
\includegraphics[width=0.37\textwidth]{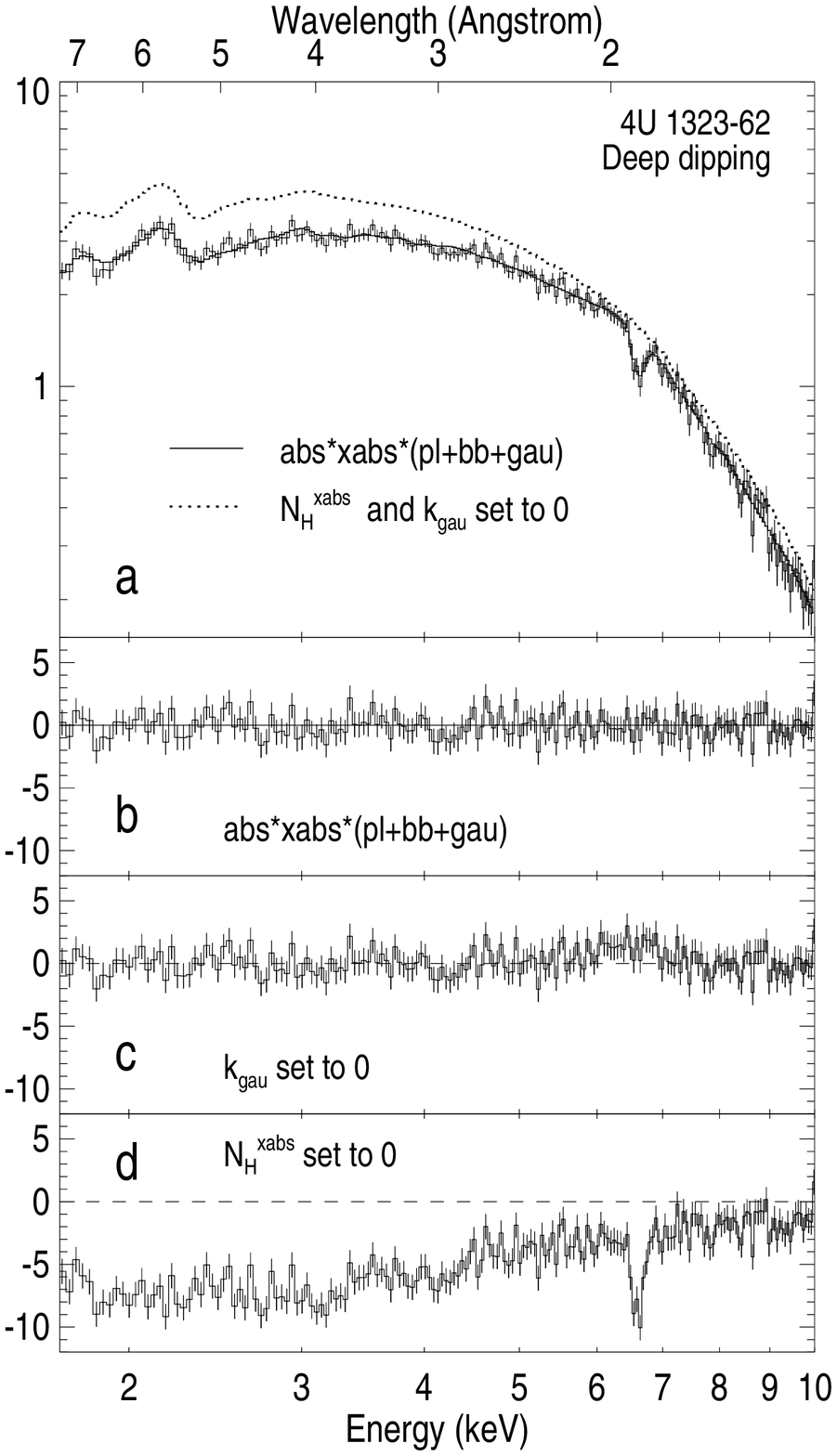}}
\caption{ {\bfseries a)} EPIC PN spectra of the \src\ persistent
  (left), shallow dipping (middle) and deep dipping (right) intervals
  fit with a model consisting of a power-law ({\tt pl}), a blackbody
  ({\tt bb}) and a broad Gaussian emission line ({\tt gau}), modified
  by absorption from neutral ({\tt abs}) and ionized ({\tt xabs})
  material (see Table~\ref{tab:spectrum}). The dotted line shows the
  model when the normalization of the Gaussian line and the column
  density of the ionized absorber are set to 0. {\bfseries b)}
  Residuals in units of standard deviations from the above model.
  {\bfseries c)} Residuals when the normalization of the Gaussian
  emission line is set to 0.  {\bfseries d)} Residuals when the column
  density of the ionized absorber is set to 0. The ionized absorber
  not only produces the narrow features near 7~keV, but also continuum
  absorption, especially during deep dipping. The fits were performed
  to combined PN and RGS spectra, but only the PN spectra are shown
  for clarity.}
\label{fig:spectrum}
\end{figure*}

Seven X-ray bursts are observed. Two of them occur during dipping,
near cycles 3.9 and 4.9. Five dips are visible.  Fig.~\ref{fig:zoomlc}
shows an expanded view of the second and third bursts with a time
resolution of 1~s and of two dips with a time resolution of 8~s.  The
dip shape varies from dip to dip.  However, the dips seem to follow a
repetitive pattern. The intensity first decreases while the hardness
ratio remains more or less constant, and at some point, both the
intensity and the hardness ratio become highly variable on timescales
between tens of seconds and minutes. The deepest segments of the dips
are clearly associated with the strongest hardening
(Fig.~\ref{fig:hrint}). We refer to the first and second part of this
pattern as ``shallow'' and ``deep'' dipping, respectively, and select
intervals corresponding to persistent, shallow dipping and deep
dipping emission as indicated in Table~\ref{tab:timeselection}.  The
0.6--10~keV persistent intensity (outside the dips and the bursts) of
\src\ increases slowly throughout the observation from $\sim$20 to
$\sim$32~\countsec\ and then decreases at the end
(Fig.~\ref{fig:lc}b). The hardness ratio of the persistent emission
increases in correlation with the intensity from $\sim$2.7 to
$\sim$3.3 (Fig.~\ref{fig:hrint}). The burst frequency seems to
increase with the intensity of the persistent emission. A
time-resolved spectral analysis indicates that none of the bursts show
evidence for photospheric radius expansion.

\subsection{Persistent, shallow and deep dipping  spectra}
\label{sec:averaged}

We extracted EPIC PN and RGS spectra for each category of emission
(Table~\ref{tab:timeselection}), excluding X-ray bursts.  Spectral
analysis was performed on combined PN and RGS spectra, as described in
Sect.~\ref{sec:spectralanalysis}.  A blackbody plus power-law model
modified by neutral absorption ({\tt abs*(pl+bb)} model within SPEX)
fits the overall continuum reasonably well in all three cases.
Examination of the spectral residuals reveals narrow absorption
features around 7~keV, superposed on a broader emission feature.  To
account for this complexity, we included a broad Gaussian emission
feature ({\tt gau}) and absorption from a photo-ionized plasma (\xabs,
see Sect.~\ref{sec:narrow}) in the model.  The best-fit parameters of
the {\tt abs*xabs*(pl+bb+gau}) are given in Table~\ref{tab:spectrum}
and the ionized absorber further described in Appendix~A.
Figs.~\ref{fig:spectrum} and \ref{fig:zoomresiduals} show the best-fit
model and residuals. The 7~keV features are very well modeled.

In all three cases, the continuum can be well described by a power-law
with a photon index of $\sim$1.9, a blackbody with a temperature,
\ktbb, of $\sim$1~keV and a Gaussian emission feature centered on
$\sim$6.6~keV with a $FWHM$ of 1.1--2~keV. The contribution of the
power-law to the total 0.5--10~keV luminosity is $\approxgt$88\%.  The
broad Gaussian emission feature is detected at a confidence level of
8.3\sig, 5\sig\ and 2.5\sig, in the persistent, shallow and deep
dipping spectrum, respectively.  The hydrogen column density, \nhabs,
is 3.5--3.9~\ttnh. The parameters describing the persistent, shallow
dipping and deep dipping continua are consistent with each other at
the 90\% confidence level.

%
%
\begin{figure}[ht]
\centerline{\hspace{1cm}\vspace{0.4cm}}
\centerline{\hspace{-0.4cm}\includegraphics[width=0.53\textwidth]{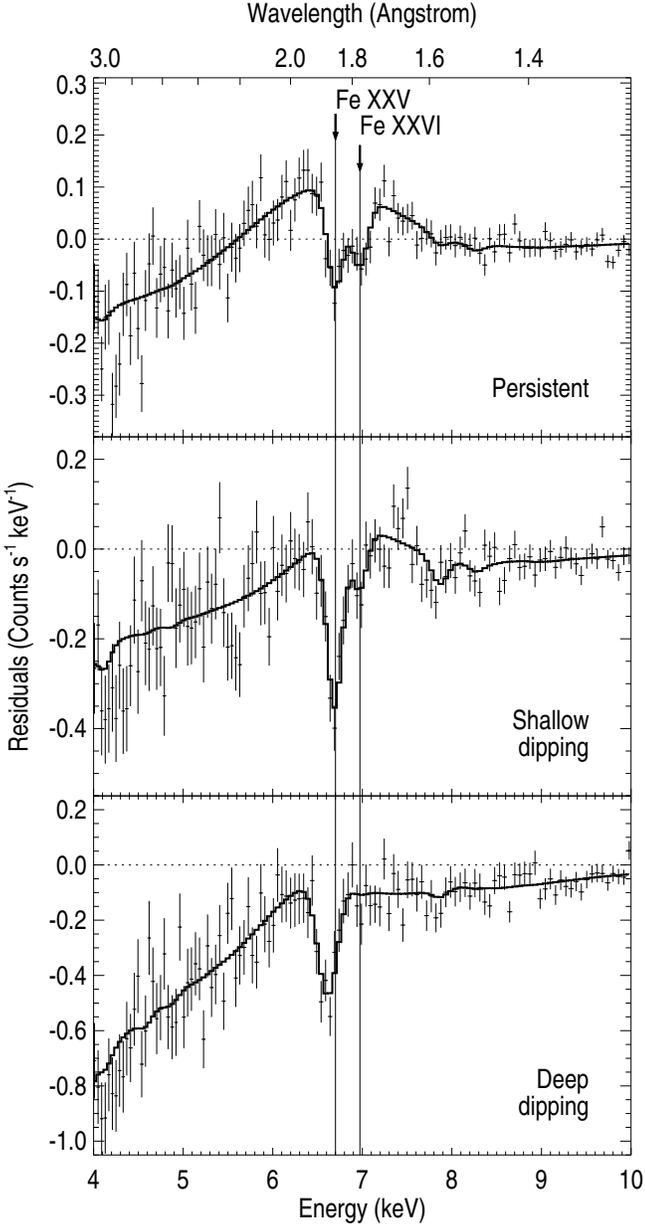}}
\caption{
  4--10 keV spectral residuals from the best-fit {\tt
  abs*xabs*(pl+bb+gau)} model (see Table~\ref{tab:spectrum}) when the
  normalization of the Gaussian emission feature, \kgau, and the
  column density of the ionized plasma, \nhxabs, are set to zero.  The
  thick line is the best-fit model minus the best-fit model when
  \kgau\ and \nhxabs\ are set to zero.  The theoretical energies of
  the \fetfive\ and \fetsix\ absorption lines are indicated.}
\label{fig:zoomresiduals}
\end{figure}

%
%
\begin{table}[!ht]
\begin{center}
\caption[]{
  Best-fits to the RGS and EPIC PN spectra of the persistent, shallow
  dipping and deep dipping emission using the {\tt
  abs*xabs*(pl+bb+gau)} model (see Sect.~\ref{sec:averaged}).  \flux\
  is the 0.5--10~keV absorbed flux. \kpl, \kbb\ and \kgau\ are the
  normalizations of the power-law (at 1~keV), blackbody and emission
  Gaussian line, respectively. \lum\ is the 0.5--10~keV unabsorbed
  luminosity assuming a source distance of 10~kpc. The contribution of
  each additive component to the total luminosity is indicated.}
\begin{tabular}{ll@{\extracolsep{-0.1cm}}c@{\extracolsep{0.15cm}}c@{\extracolsep{0.15cm}}c}

\hline
\hline
\noalign {\smallskip}
\multicolumn{2}{l}{} Component   & Pers.  & \multicolumn{2}{c}{Dipping}\\
\multicolumn{2}{l}{Parameter}  & & Shallow & Deep \\
\noalign {\smallskip}
\hline
\noalign {\smallskip}
\noalign {\smallskip}
\multicolumn{2}{r}{\tt abs}  & & & \\
\multicolumn{2}{l}{\nhabs\ {\small($10^{22}$ cm$^{-2}$)}}       &  3.5 $^{+0.1}_{-0.2}$ & 3.6 $^{+0.2}_{-0.3}$  & 3.9 $^{+0.4}_{-0.6}$\\
\noalign {\smallskip}
\multicolumn{2}{r}{\xabs}  & & & \\
\multicolumn{2}{l}{ \nhxabs\ {\small($10^{22}$ cm$^{-2}$)}}     &  3.6 $^{+1.0}_{-0.9}$ & 6 $^{+5}_{-2}$        & 14 $^{+10}_{-7}$\\
\multicolumn{2}{l}{\logxi\ {\small(\xiunit)}}                   & 3.90 $^{+0.08}_{-0.09}$ & 3.6 $^{+0.1}_{-0.2}$        & 3.0 $\pm~0.2$ \\
\multicolumn{2}{l}{ \sigmav\ {\small(km s$^{-1}$)}}     & 1700 $\pm$ 1000      & 600 $^{+2000}_{-400}$ & 90 $^{+80}_{-40}$ \\
\multicolumn{2}{l}{  $v$ {\small(km s$^{-1}$)} }                & -20 $^{+900}_{-1000}$ &  0 $^{+2000}_{-700}$  & 0 $^{+3000}_{-800}$ \\
\noalign {\smallskip}
\multicolumn{2}{r}{\tt pl}      &  & & \\  
\multicolumn{2}{l}{\phind}                              &  1.90 $^{+0.06}_{-0.10}$      & 1.9 $\pm$ 0.1         & 1.8 $^{+0.2}_{-0.3}$\\
\multicolumn{2}{l}{\kpl\ {\small ($10^{44}$ ph. s$^{-1}$ keV$^{-1}$)}}  & 9 $^{+1}_{-2}$        & 10 $\pm$ 2            & 10 $^{+4}_{-3}$\\
\noalign {\smallskip}
\noalign {\smallskip}
\multicolumn{2}{r}{\tt bb}    & & & \\  
\multicolumn{2}{l}{\ktbb\ {\small(keV)}}                        & 0.98 $^{+0.06}_{-0.06}$       & 1.1 $\pm$ 0.1         & 1.2 $^{+0.4}_{-0.2}$\\
\multicolumn{2}{l}{\kbb\  {\small($10^{11}$ cm$^{2}$)}} & 6  $^{+4}_{-2}$       &  4 $^{+3}_{-2}$               & 2 $^{+6}_{-1}$\\
\noalign {\smallskip}
\noalign {\smallskip}
\multicolumn{2}{r}{\tt gau}    & & & \\  
\multicolumn{2}{l}{ \egau\ {\small(keV)}}               & 6.6 $^{+0.1}_{-0.2}$          & 6.9 $\pm$ 0.2                 &  6.4 $^{+0.2}_{-0.3}$\\
\multicolumn{2}{l}{ $FWHM$ {\small(keV)}}               &  2.0 $^{+0.6}_{-0.4}$ &  1.3 $^{+1.0}_{-0.5}$         & 1.1 $^{+1.0}_{-0.5}$\\
\multicolumn{2}{l}{ \kgau\ {\small(10$^{42}$ ph s$^{-1}$)}} & 5 $^{+3}_{-1}$ & 3 $^{+2}_{-1}$   & 3 $^{+4}_{-2}$ \\
\noalign {\smallskip}
\hline

\noalign {\smallskip}
        \multicolumn{2}{l}{\flux\ {\small (10$^{-10}$ \ergcms)}} & 1.97 &  1.76  & 1.27 \\
        \multicolumn{2}{l}{\lum\ {\small (10$^{36}$ \ergs)}} & 5.19  & 4.75  & 3.90  \\
        \multicolumn{2}{l}{$L_{\tt pl}$/\lum\  (\%)} & 88.3 &  87.5 &  87.3 \\
        \multicolumn{2}{l}{$L_{\tt bb}$/\lum\  (\%)} & 10.7 &  11.9 & 11.9  \\
        \multicolumn{2}{l}{$L_{\tt gau}$/\lum\  (\%)} & 1.0 &  0.6 &  0.8 \\
\noalign {\smallskip}
\hline
\noalign {\smallskip}
        \multicolumn{2}{l}{\rchisq} & 1.13 & 1.19 & 0.95 \\
        \multicolumn{2}{l}{d.o.f.} & 315 & 235 & 207 \\
        \multicolumn{2}{l}{Exposure (ks)} & 30.4 & 10.8 & 4.2 \\
\noalign {\smallskip}
\hline
\end{tabular}
\label{tab:spectrum}
\end{center}
\end{table}
 
In contrast, the properties of the ionized absorber show an evolution
from persistent to deep dipping (except for $v$ which is poorly
constrained due to the limited energy resolution).  The hydrogen
column density of the ionized absorber, \nhxabs, increases marginally
from ($3.6\,^{+1.0}_{-0.9}$)~\ttnh\ to ($14\,^{+10}_{-7}$)~\ttnh,
while the photo-ionization parameter decreases significantly from
persistent, with \logxi\ of $3.90\, ^{+0.08} _{-0.09}$, to deep
dipping with \logxi\ of $3.0 \pm 0.2$.  The strongest lines in the
persistent spectrum are the \fetfive\ and
\fetsix\ 1s-2p ones (see Fig.~\ref{fig:zoomresiduals} and
Table~\ref{tab:xabslines}).  During shallow dipping, the
\ew\ of the \fetfive\ line increases while that of \fetsix\
decreases. In the deep dipping stage, the \fetsix\ line has become
weak, and the observed absorption feature at 6.60~keV is due to a
blend of \fetfive\ to \fetone\ lines.  Therefore, the spectral
differences in the narrow absorption features between persistent and
dipping emission can be modeled by the presence of an ionized absorber
with a higher column density and lower ionization parameter during
dipping intervals.

%
%
\begin{figure}[!ht]
\centerline{\hfill \includegraphics[width=0.5\textwidth]{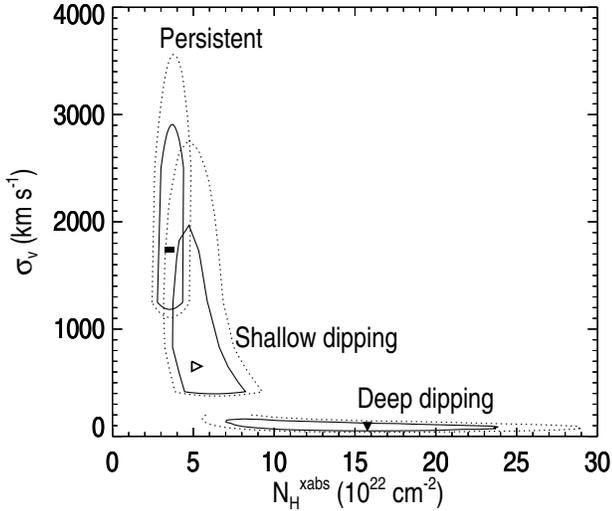} \hfill}
\caption{Confidence contours of \sigmav\ versus \nhxabs\ for the
    persistent, shallow dipping and deep dipping spectra fit with the
    model described in Table~\ref{tab:spectrum}. The continuous and
    dotted lines indicate the 68.3\% and 90\% confidence levels,
    respectively, for two varying parameters.  In the persistent
    spectrum, \nhxabs\ is well constrained whereas \sigmav\ is poorly
    constrained.  This is evidence for the absorption lines being
    unsaturated. In the deep dipping spectrum, the opposite trend is
    observed, indicating that the lines are saturated.}
\label{fig:contour}
\end{figure}

\sigmav\ decreases from $1700 \pm 1000$ to
$90\,^{+80}_{-40}$~\kms, from persistent to deep dipping.
Fig.~\ref{fig:contour} shows the confidence contours of \sigmav\
versus \nhxabs\ for the persistent, shallow dipping and deep dipping
spectra. In the persistent spectrum, \nhxabs\ is well constrained
whereas \sigmav\ is poorly constrained.  This is evidence for the
absorption lines being unsaturated, i.e. on the linear part of the
curve of growth, where their $EW$s increase linearly with the column
density, independently of the velocity broadening. On the contrary, in
the deep dipping spectrum, \nhxabs\ is poorly constrained whereas
\sigmav\ is well constrained. This indicates that the lines are
saturated: their $EW$s do not strongly depend on the column density
anymore, but rather increase with the velocity broadening.

\begin{figure*}[!ht]
\centerline{\hspace{1cm}\vspace{0.4cm}}
\centerline{\hfill\includegraphics[angle=0,width=0.4\textwidth]{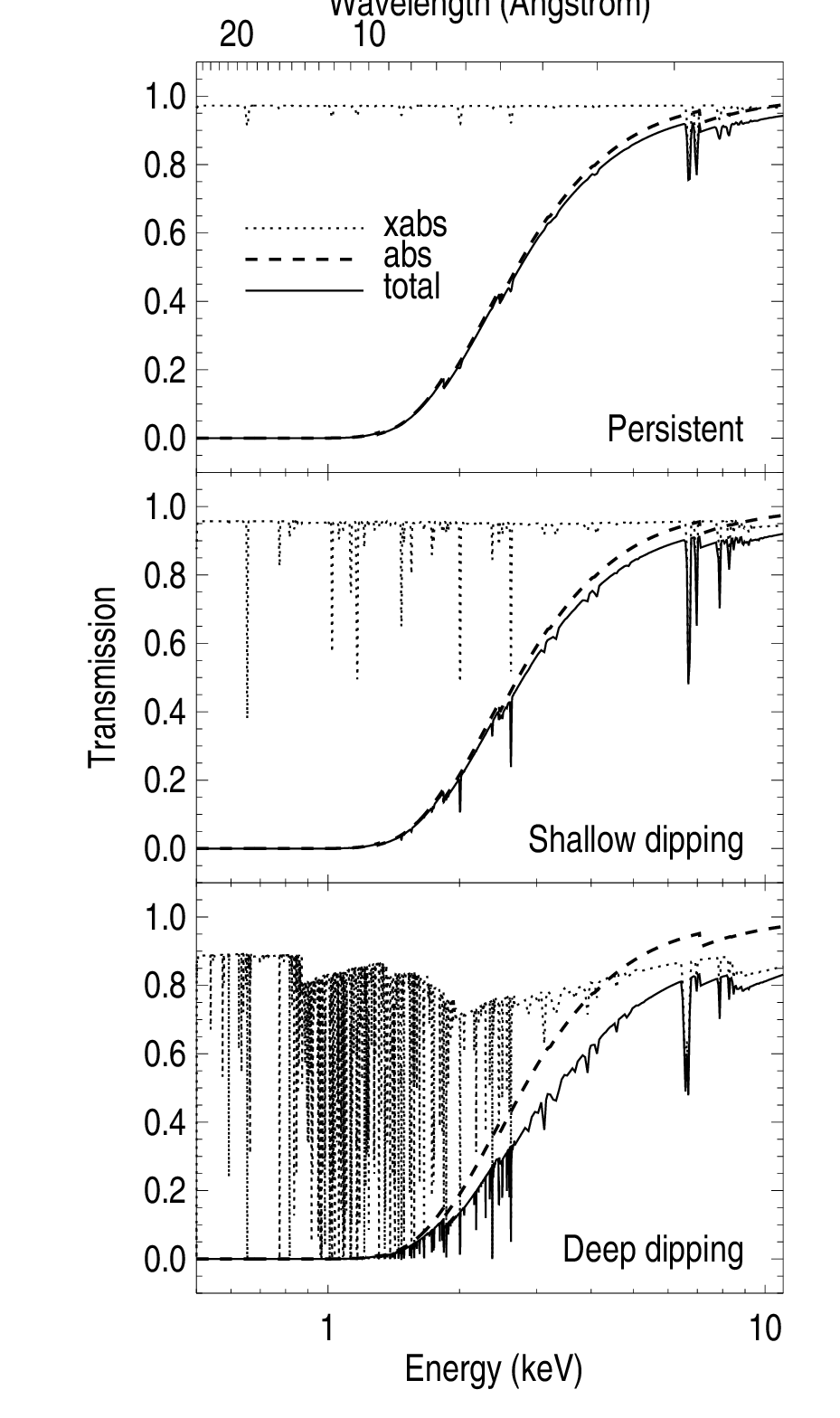}\hfill\includegraphics[angle=0,width=0.4\textwidth]{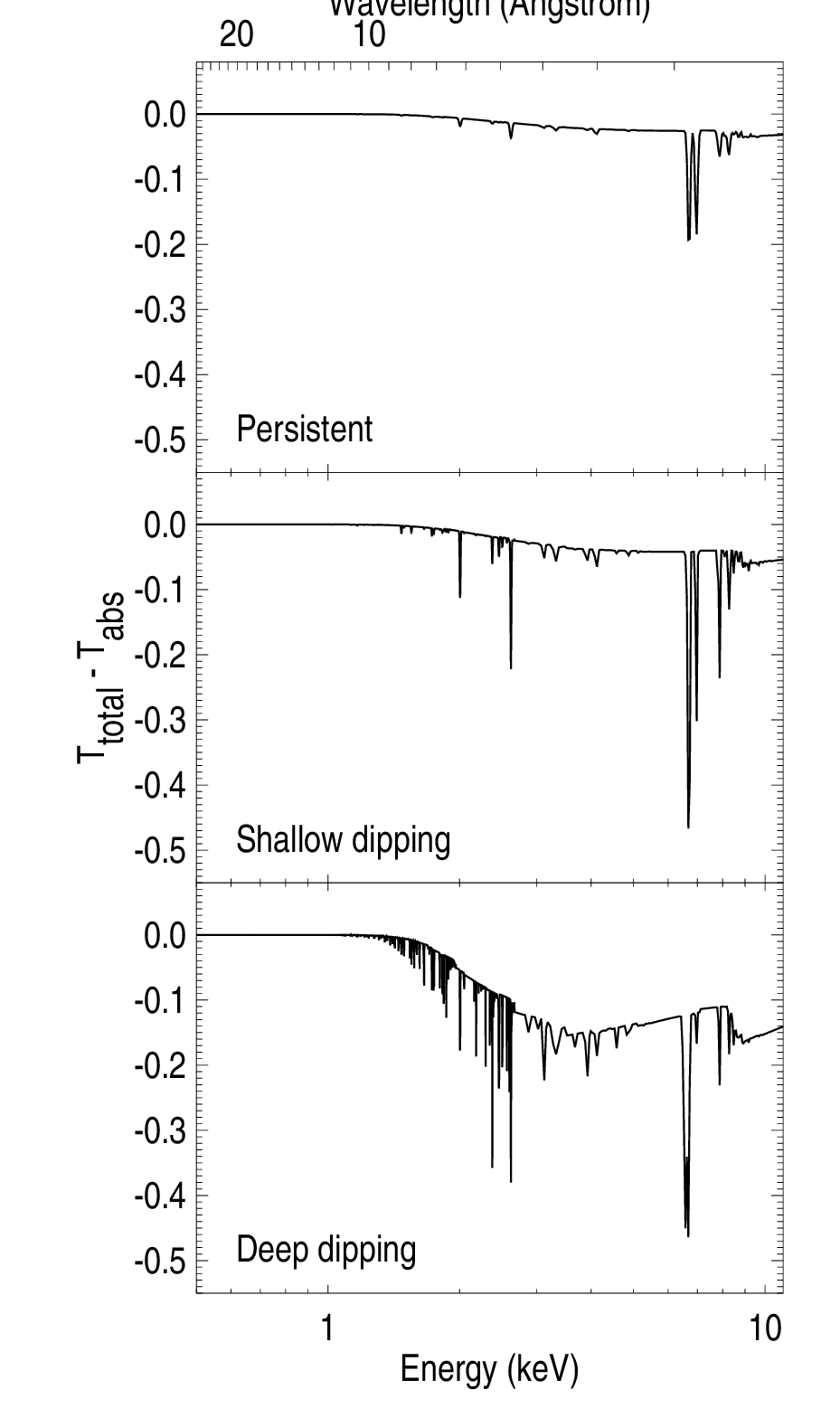}\hfill}
\caption{
  The left panels show the transmission of the ionized (dotted line,
  \xabs) and neutral absorbers (dashed line, {\tt abs}), and the total
  transmission (thick line) for the persistent, shallow and deep
  dipping spectra using the {\tt abs*xabs*(pl+bb+gau)} model
  (Table~\ref{tab:spectrum}). The right panels show the difference
  between the total transmission and the transmission of the neutral
  absorber to emphasize the contribution of the ionized absorber and
  its increasing energy dependence with dipping.}
\label{fig:transmission}
\end{figure*}

Fig.~\ref{fig:transmission} shows the transmission of the ionized and
neutral absorbers individually, and the total transmission for the
persistent, shallow and deep dipping spectra fit with the model
described in Table~\ref{tab:spectrum}.  The transmission of the
ionized absorber is nearly unity for the persistent emission. Only the
\fetfive\ and \fetsix\ lines are strongly evident.  The number of
lines increases from the persistent to the deep dipping stage.  In
addition, due to the continuum opacity of the ions, the continuum
transmission becomes highly affected, particularly between 1.5--4~keV.
This effect contributes strongly to the change in spectral ratio
between the persistent and dipping intervals.

Table~\ref{tab:gaussians} shows the best-fit parameters of the most
prominent absorption features fit with simple Gaussian profiles.
Table~\ref{tab:upperlimits} shows the upper limits on the \ew\ and
optical depth, $\tau$, of non detected absorption lines and edges,
respectively. The measured values and upper limits are all consistent
with the values predicted by the photo-ionized absorber model
(Tables~\ref{tab:xabslines} and \ref{tab:xabsedges}). Note
that, because of the limited spectral resolution, the measured \ews\
in Table~\ref{tab:gaussians} should be compared to the sum of the
\ews\ of the various lines in Table~\ref{tab:xabslines} likely to
contribute to the measured feature, i.e. the various lines whose
energy is within the width of the measured feature.

%
%
\begin{table}[!ht]
\begin{center}
\caption[]{Best-fit parameters of the Gaussian 
profiles fit to the most prominent absorption features in the
persistent, shallow and deep dipping average spectra.}
\begin{tabular}{lccc}

\hline
\hline
\noalign {\smallskip}
	 & Persistent	& \multicolumn{2}{c}{Dipping} \\
	& 		& Shallow 	& Deep \\
\noalign {\smallskip}
\hline
\noalign {\smallskip}
Ident.		& \fetfive\		& \fetfive\		& {\bfseries \ion{Fe}{xxi--xxv}} \\
$E$ (keV) 	& 6.68 $\pm$ 0.04 		& 6.68 $\pm$ 0.03		& 6.60 $\pm$ 0.02 \\
\sig\ (eV)		& $<$100			& $<$75  			&$<$120 \\
\ew\   (eV)	& $25\,_{-7}^{+19}$ 	& $53\,_{-12}^{+17} $  	& $65\,_{-21}^{+63} $  \\
\noalign {\smallskip}
\noalign {\smallskip}
Ident.		& \fetsix\  		&  			& \\
$E$ 		& 6.97 $\pm~0.05$		& 			& \\ 
\sig\ 		& $<$136			& 			&  \\
\ew\ 		& $24\,_{-7}^{+21} $	&    			&	\\
\noalign {\smallskip}
\hline
\end{tabular}
\label{tab:gaussians}
\end{center}
\end{table}

%
%
\begin{table}[!h]
\begin{center}
\caption[]{
Upper limits on the \ew\ and optical depth, $\tau$, of non detected
absorption lines and edges, respectively, predicted by the
photo-ionization model (see Tables~\ref{tab:xabslines} and
\ref{tab:xabsedges}). We derive upper limit on the \ew\ of a line 
by including a Gaussian profile with a width fixed to 0 at the
required energy.}
\begin{tabularx}{\linewidth}{llXXX}

\hline
\hline
\noalign {\smallskip}
Ident. & Energy  & Persistent & \multicolumn{2}{c}{Dipping} \\
& (keV) & & Shallow & Deep \\
\noalign {\smallskip}
\hline
\noalign {\smallskip}
\multicolumn{2}{c}{Lines}	& \multicolumn{3}{c}{\ew\ (eV)} \\
\noalign {\smallskip}
\ssixteen\		& 2.623 	& $<$4	&    $<$7		& $<$8	\\
\fetsix\		& 6.973 	& 	& 21 $\pm~12$	& $<$34 \\
\fetfive\  		& 7.881 	& $<$8	&    $<$25	& $<$38	\\
\fetfive\  		& 8.296 	& $<$8	&    $<$17 	& $<$25	\\
\noalign {\smallskip}
\hline
\noalign {\smallskip}
\multicolumn{2}{c}{Edges}	& \multicolumn{3}{c}{$\tau$} \\
\noalign {\smallskip}
\sifourteen\ 	& 2.67	&		&  		& $<$0.10 \\
\fetfive\ 		& 8.83	& $<$0.02	& $<$0.04	& $<$0.06 \\
\noalign {\smallskip}
\hline
\end{tabularx}
\label{tab:upperlimits}
\end{center}
\end{table}

%
%

There are fainter features remaining in the spectra. An absorption
feature is visible in the persistent spectrum at $4.25 \, ^{+0.11}
_{-0.07}$~keV and can be fit by a Gaussian with a width $\sigma
<$330~eV and and \ew\ of $6\,_{-2}^{+16}$~eV.  A weak absorption line
is predicted by \xabs\ at 4.108~keV due to \ion{Ca}{xx}
(Table~\ref{tab:xabslines}). However, this energy is inconsistent at a
3.3\sig\ level with that of the observed feature.  Alternatively, a
dramatic blue-shift of $~$9700 km~s$^{-1}$ would be required for the
observed feature to be interpreted as \ion{Ca}{xx}. This is
incompatible with the blue-shift of $<$1020~km~s$^{-1}$ determined
primarily from the strong \fetfive\ and \fetsix\ absorption lines
during persistent emission (Table~\ref{tab:spectrum}). A 4.2~keV
feature is also visible in the shallow and deep dipping spectra, and
can be described by a Gaussian profile with parameters consistent with
those obtained from the persistent spectrum. A broad ``bump'' near
3.5~keV is also present in all three spectra.  Its origin is unclear,
but it may be due to incorrect modeling of the nearby Au mirror edges
connected with the structured residuals near $\sim$2.2~keV and
$\sim$2.8~keV.

\subsubsection{The broad Fe emission line}

In order to investigate whether the large width ($FWHM = \bf
2.0\,^{+0.6}_{-0.4}$~keV, see Table~\ref{tab:spectrum}) of the Gaussian emission
feature detected at 6.6~keV could be due to relativistic effects, we
modeled the line in the persistent spectrum using the
\citet{laor91apj} model. The resulting fit is acceptable (\rchisq\ of
1.16 for 311 d.o.f.), but not statistically better than the one
obtained using a simple broad Gaussian line. Most of the parameters of
the Laor model are poorly constrained. The Gaussian line energy is
$6.6\,^{+0.1}_{-0.2}$~keV, and the normalization is $5\,
^{+22}_{-1}$ 10$^{\bf 42}$ ph s$^{-1}$. The derived inner radius for
the emitting region is $11\, ^{+6}_{-10}$~G$M$/c$^2$ and the outer
radius is $>$92~G$M$/c$^2$. The emissivity slope is $3\, ^{+2}_{-1}$,
the emissivity scale $<$69 and the inclination
$87.5\degmark\,^{+2.5}_{-1.3}$.  Thus, while the large width may be
due to relativistic broadening, we cannot exclude that other
broadening mechanisms such as Compton scattering and emission from a
range of ionization states contribute to the broadening.

\subsection{Time-resolved spectral fits}
\label{sec:timeresolved}

The results of Sect.~\ref{sec:averaged} suggest that, to first order,
the only change in the spectrum during dips is the appearance of an
extra absorption component that affects the underlying ``persistent''
spectrum. A closer inspection of Figs.~\ref{fig:lc} and
\ref{fig:hrint} shows, however, that the persistent flux is also
changing on a long time scale and that the persistent hardness ratio
is evolving in correlation with the flux. These secular spectral
changes were not accounted for in Sect.~\ref{sec:averaged}.  In this
section, we want to test the hypothesis that the spectral changes
observed between a persistent and a dipping segment can be explained
uniquely by variations in the properties of the absorbers in the
line-of-sight, as long as any secular spectral change attributed to
the underlying source is taken into account.

%
%
\begin{figure*}[!ht]
\centerline{\hfill
\includegraphics[width=0.5\textwidth]{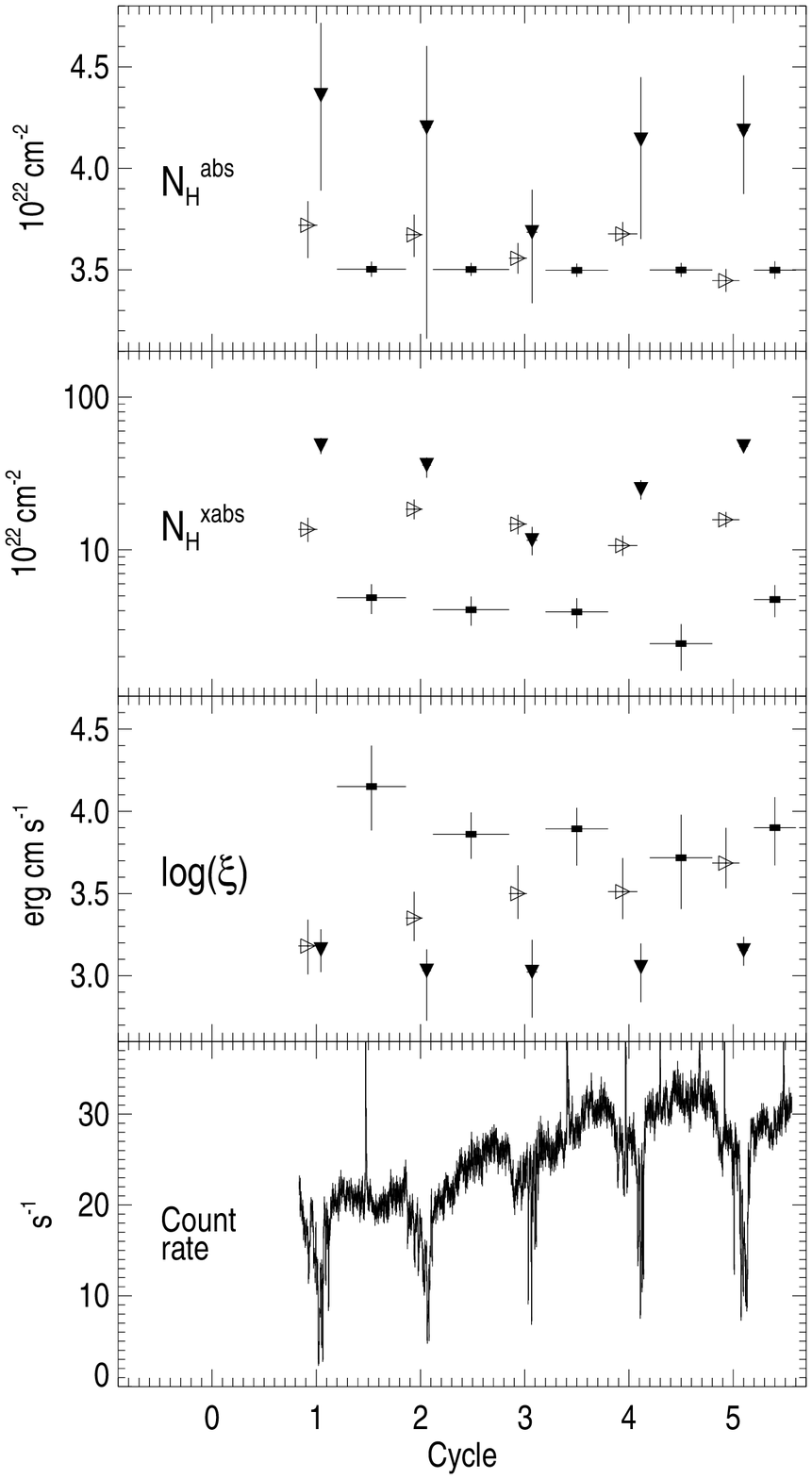}
\hfill
\includegraphics[width=0.5\textwidth]{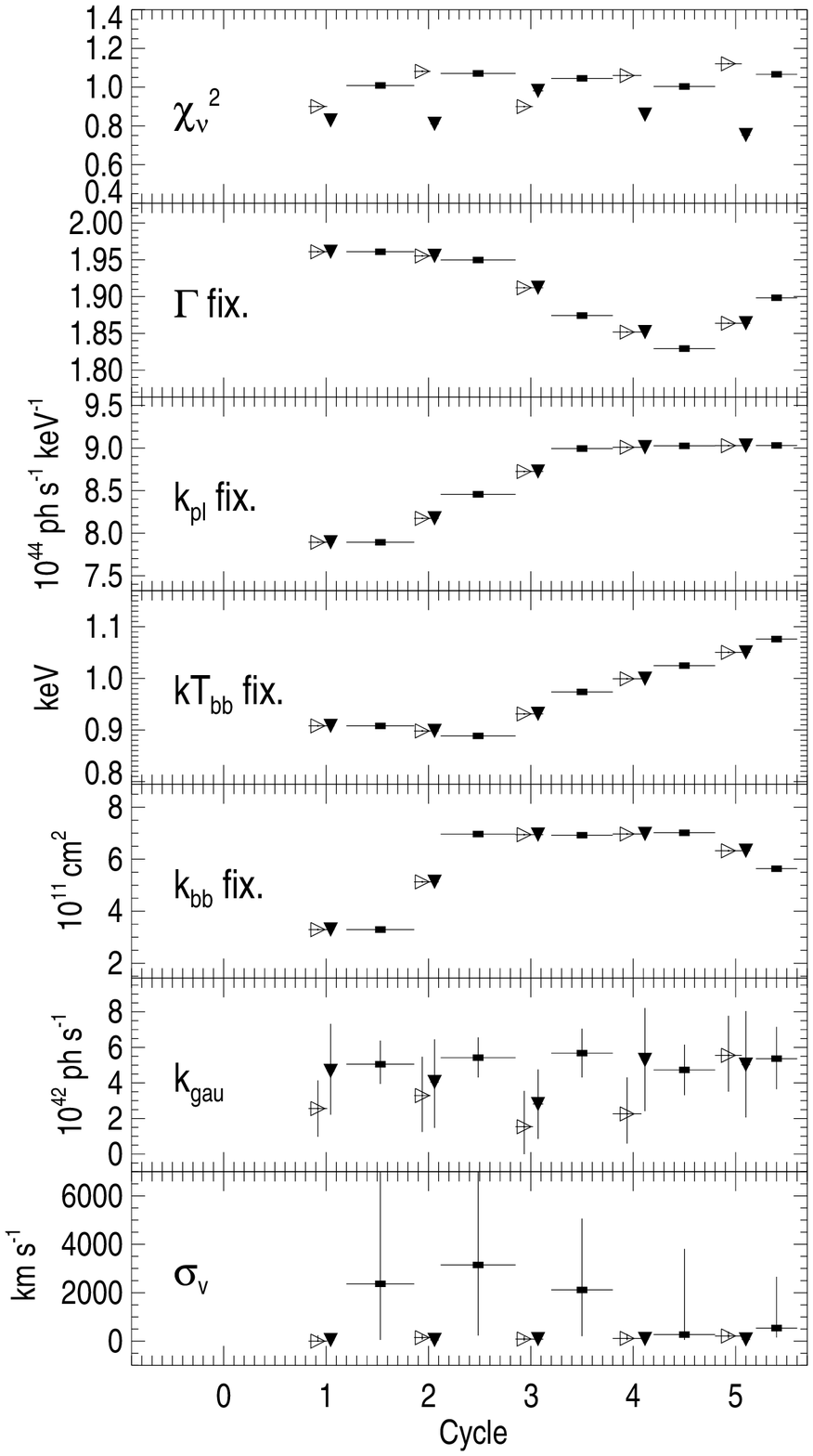}
\hfill}
\caption{
  Best-fit parameters of the individual spectra using the {\tt
    abs*xabs*(pl+bb+gau)} model and fixing \phind, \kpl, \ktbb\ and
  \kbb\ to their persistent-interpolated values (see text). Filled
  rectangles, and empty and filled triangles indicate the persistent,
  shallow and deep dipping intervals, respectively.  The parameters
  are defined in Sects.~\ref{sec:narrow} and~\ref{sec:averaged}.
  ``fix.'' indicates that a parameter was fixed at the value shown in
  the figure. The parameters that account for the spectral changes
  between persistent and dipping emission are shown in the 3 upper
  left panels.  The top right panel shows the \rchisq\ of the fits.
  The bottom left panel shows the 0.6--10~keV PN lightcurve. }
\label{fig:parameters}
\end{figure*}

To account for these secular changes in the persistent emission, we
assumed that the neutral absorption remains constant, and we fixed it
to our best-fit value of the average persistent spectrum
(Table~\ref{tab:spectrum}), \nhabs~=~3.5~\ttnh, that we call the
``galactic'' value, although part of the neutral absorber might be
located close to the binary system. We then fit the individual
persistent spectra (Table~\ref{tab:timeselection}) with the {\tt
abs*xabs*(pl+bb+gau)} model fixing \nhabs\ to the galactic value.  We
also fixed $v$, \sigmav\ and the parameters of the Gaussian emission
line to the values obtained for the average persistent spectrum
(Table~\ref{tab:spectrum}). Thus, the only parameters free to vary
were \phind, \kpl, \ktbb, \kbb, \nhxabs\ and $\xi$.  The fits are
acceptable, and the values of \phind, \kpl, \ktbb\ and \kbb\ obtained
in this way for each of the individual persistent spectra are
considered as our ``reference values''.

Finally we fit all the individual spectra (persistent, shallow and
deep dipping) using the {\tt abs*xabs*(pl+bb+gau}) model.  \nhabs\ was
allowed to vary, but \phind, \kpl, \ktbb\ and \kbb\ were fixed to the
values obtained directly from the previous fits (the reference values)
in the case of the persistent spectra, and to values interpolated from
adjacent reference values in the case of the dipping spectra.  Since
the XMM-Newton observation started while \src\ was probably dipping,
the interpolated values for the first dip were taken equal to the
reference values of the following persistent segment. $v$ was kept
fixed to 0 for both the persistent and dipping segments. The energy
and $FWHM$ of the Gaussian emission line were fixed to the values
obtained for the corresponding average spectra
(Table~\ref{tab:spectrum}), but its normalization was left free to
vary.

%
%
\begin{table}[!ht]
\begin{center}
\caption[]{Weighted averages of \nhabs, \nhxabs\ and \logxi, 
during persistent, shallow and deep dipping emission, using the
results from the individual spectral fits
(Fig.~\ref{fig:parameters}). We did not take into account the results
from the ``deep'' dipping interval near cycle~3 which resembles a
shallow dipping interval.}
\begin{tabular}{ll@{\extracolsep{-0.1cm}}ccc}
\hline
\hline
\noalign {\smallskip}
\multicolumn{2}{r}{}   & Persistent  & \multicolumn{2}{c}{Dipping}\\
\multicolumn{2}{l}{}  & & Shallow & Deep \\
\noalign {\smallskip}
\hline
\noalign {\smallskip}
\multicolumn{2}{l}{\nhabs\ {\small($10^{22}$ cm$^{-2}$)}}       &   $3.50 \pm 0.02$             & $3.58 \pm 0.03$       &  $4.2 \pm 0.2$ \\
\noalign {\smallskip}
\multicolumn{2}{l}{ \nhxabs\ {\small($10^{22}$ cm$^{-2}$)}}     &   $3.8 \pm 0.4$       &  $14 \pm 1$   & $ 37 \pm 2$   \\
\noalign {\smallskip}
\multicolumn{2}{l}{\logxi}                      &  $3.9 \pm 0.1$      &  $3.43 \pm 0.08$      &  $3.13 \pm 0.07$ \\
\noalign {\smallskip}
\hline
\end{tabular}
\label{tab:weightedaverages}
\end{center}
\end{table}
 
The results of this set of fits are shown in
Fig.~\ref{fig:parameters}. The fit quality remains acceptable with
\rchisq\ between 0.75--1.13 for 153 to 184 d.o.f. 
In the individual segments, the values of \kgau, the normalization of
the Gaussian emission line, and of \sigmav\ are all found to be
consistent with the values obtained from the average spectra.

The most interesting results concern \nhabs, \nhxabs\ and $\xi$, and
are shown in the upper left panels of Fig.~\ref{fig:parameters}. For
these parameters, the values obtained for the persistent and dipping
segments form distinguishable and well constrained groups. The \nhabs\
values for all persistent intervals are consistent with each other and
have a weighted average of ($3.50 \pm 0.02$)~\ttnh\
(Table~\ref{tab:weightedaverages}) in agreement with the galactic
value, as expected. \nhabs\ is significantly higher in the deep
dipping intervals.  The only exception comes from the deep dipping
segment around cycle~3; this is probably due to the dipping activity
being very deep for a short time, and resembling shallow dipping
elsewhere (see Fig.~\ref{fig:zoomlc}). If we exclude this segment, the
weighted average value of \nhabs\ during deep dipping intervals is
($4.2
\pm 0.2$)~\ttnh\ (Table~\ref{tab:weightedaverages}). This is larger
than the persistent value at a 5.7\sig\ level, and represents an
increase by a factor 1.2. In the shallow dipping intervals, \nhabs\ is
intermediate between the galactic and the deep dipping values.

\nhxabs\ shows the same behavior as \nhabs, but the difference between
the persistent and the dipping values is even more pronounced (note
the logarithmic scale for this parameter in
Fig.~\ref{fig:parameters}). The weighted averages for each emission
category are listed in Table~\ref{tab:weightedaverages}.  \nhxabs\ is
($3.8 \pm 0.4$)~\ttnh\ during persistent segments and ($37 \pm
2$)~\ttnh, a factor $\sim$10 higher, during deep dipping
intervals. \nhxabs\ is intermediate during the shallow dipping and
seems to increase with the count rate.  The \nhxabs\ values obtained
from the time-resolved spectral fits are significantly larger than
those obtained from the average spectra. This is not unexpected and
actually constitutes the main result of this time-resolved
analysis. Indeed, in the average spectral analysis, all the parameters
were free to vary, and spectral changes could be accounted for by a
simultaneous moderate change of \nhxabs\ together with all the other
parameters, including the normalizations of the additive
components. On the contrary, in the time-resolved spectral analysis,
the parameters of the additive components were fixed to their
underlying persistent values, and the spectral changes from persistent
to dipping intervals were forced to be accounted for only by larger
amounts of absorbing material.

Finally, the values of $\xi$ can also be clearly distinguished between
persistent and dipping segments. On average,
\logxi\ is $3.9 \pm 0.1$ during persistent intervals, and only
$3.13 \pm 0.07$ during deep dipping, which is lower at a 10\sig\
level.  The values for the shallow dipping intervals are intermediate
(Table~\ref{tab:weightedaverages}).

Summarizing, the spectral changes in both the absorption narrow
features {\it and} the continuum observed between persistent and
dipping emission can be modeled simply by variations in the properties
of the neutral and ionized absorbers, with the ionized absorber
playing the main role.  During dipping, both the amount of neutral and
ionized absorbing material in the line-of-sight is higher with a much
larger difference for the ionized material. In addition, the material
in the line-of-sight is less ionized during dipping than during
persistent intervals.

\section{Discussion}

We report the detection of narrow \fetfive\ (He-like) and \fetsix\
(H-like) 1s-2p resonant X-ray absorption lines at 6.68$ \pm 0.04$ and
$6.97 \pm 0.05$~keV in the persistent emission of \src.  These
features are superposed on a broad emission feature centered on
$6.6\,^{+0.1}_{-0.2}$~keV. During dipping intervals the $EW$ of the
\fetfive\ feature increases while that of the \fetsix\ feature
decreases implying the presence of less strongly ionized material in
the line-of-sight during dips. The overall 1.0--10~keV spectrum
becomes harder during the dips, but this change is inconsistent with a
simple increase in absorption by cool material. We demonstrate that
both the changes in the continuum {\it and} absorption lines observed
between persistent and dipping intervals can be modeled mainly by
changes in the properties of an ionized absorber which has a higher
column density and lower ionization parameter during dipping intervals
than during persistent intervals.

\fetfive\ or \fetsix\ absorption lines near 7~keV were reported from
the micro-quasars \gro\ \citep{1655:ueda98apj,1655:yamaoka01pasj},
\grs\ \citep{1915:kotani00apj,1915:lee02apj} and \seventeen\
\citep[][the classification as a micro-quasar is
preliminary]{1743:miller04apj}, and from the neutron star systems
\cir\
\citep{cirx1:brandt96mnras,cirx1:brandt00apjl,cirx1:schulz02apj}, \gx\
\citep{gx13:ueda01apjl,gx13:sidoli02aa,gx13:ueda04apj}, \mxb\
\citep{1658:sidoli01aa}, \bigdip\ \citep{1624:parmar02aa}, \twelve\
\citep{1254:boirin03aa}, \nineteen\ \citep{1916:boirin04aa} and now
\src.

All these systems except \grs\ and \gx\ show dipping activity
indicating that their inclination is 60--80$\degmark$. Note that the
report of dipping activity in \seventeen\ is preliminary
\citep{1743:miller04apj}. An inclination of 70$\degmark$ is attributed
to \grs\ assuming that the superluminal jets are perpendicular to the
accretion disk \citep{1915:mirabel94nature}. The inclination of \gx\ 
is unknown, but its 24~day periodic energy-dependent X-ray modulation
\citep{gx13:corbet03apj} might indicate a relatively high inclination.
Therefore, most of the systems exhibiting \fetfive\ or \fetsix\ 
absorption lines are viewed relatively close to edge-on.  This and the
lack of any orbital phase dependence of the features
\citep[e.g.,][]{1655:yamaoka01pasj,1254:boirin03aa} except during
dips, suggest that the highly-ionized absorber is located in a thin
cylindrical geometry around the compact object.  Such highly-ionized
disk atmosphere or wind is probably a common feature of accreting
binaries, but is preferentially detected in the dipping sources,
presumably due to being viewed at inclination angles within
$\sim$$30\degmark$ of the orbital plane.

{\it Chandra} HETGS observations revealed blue-shifts of
$\sim$500~\kms\ for the lines from \cir, \gx\ and \seventeen,
interpreted as evidence for outflows in these systems
\citep{cirx1:schulz02apj,gx13:ueda04apj,1743:miller04apj}.  The
absorption lines from the other binaries do not appear to be strongly
blue-shifted. However, most of the results come from ASCA Solid-state
Imaging Spectrometer and XMM-Newton EPIC which has a factor $\sim$4
poorer energy resolution than the HETGS at Fe-K. Therefore, the
geometry and dynamics of the highly-ionized plasma remain an open
question. Whether the plasma is in- or out-flowing and its velocity
might depend on parameters such as the luminosity and the distance
between the absorber and the ionizing source.  The upper limit to any
blue-shift during persistent emission from \src\ is 1020~km~s$^{-1}$.

The emission line detected at $6.6\,^{+0.1}_{-0.2}$~keV in the
persistent spectrum of \src\ is very broad, with a $FWHM$ of
$2.0\,^{+0.6}_{-0.4}$~keV.  While relativistic broadening cannot be
excluded, the conditions for such an effect to occur are not expected
to be met in neutron star systems like \src\ as easily as in black
hole systems where relativistically broadened lines are usually
observed. Furthermore, the inclination of
$87.5\degmark\,^{+2.5}_{-1.3}$ derived from the Laor model is
inconsistent with the inclination of $<$$80\degmark$ inferred from the
absence of eclipses in \src, assuming a mass of 1.4~\msun\ and
1~\msun\ for the neutron star and the secondary star, respectively.
Alternatively, a secondary mass of $\approxlt$0.3~\msun\ would be
required to make the two inclinations consistent with each other.
Other mechanisms such as Compton scattering, rotational velocity
broadening and emission from a range of ionization states may
contribute to the broadening in \src.  We caution that the detection
and width of such an emission Gaussian depends on the continuum
modeling which is poorly constrained due to the lack of data above
10~keV. However, an emission line was reported at $6.43 \pm 0.21$~keV
in the 2--20~keV RXTE Proportional Counter Array spectrum of \src\
\citep[the line width \sig\ was fixed to
0.4~keV,][]{1323:barnard01aa}.  Furthermore, broad Fe lines have been
detected using XMM-Newton from other neutron star systems that show
highly-ionized absorption features such as \mxb\ \citep[$FWHM = 1.4 \,
^{+0.3}_{-0.4}$~keV,][]{1658:sidoli01aa} and \gx\ \citep[$FWHM = 1.9
\pm 0.5$~keV,][]{gx13:sidoli02aa}.

Changes in the \fetsix\ and \fetfive\ absorption features between
persistent and dipping intervals have been reported from \cir\
\citep{cirx1:schulz02apj}, \seventeen\ \citep{1743:miller04apj}, and
from the classical dipping LMXBs \bigdip\ \citep{1624:parmar02aa} and
\nineteen\ \citep{1916:boirin04aa}.  From persistent to dipping
intervals, the strength of the He-like feature increases while that of
the H-like feature decreases, indicating that the material in the
line-of-sight is less ionized during dipping. In the case of \mxb,
whilst \citet{1658:sidoli01aa} report that there is no obvious orbital
dependence of the $EW$s of the Fe features, their Fig.~4 shows that
the ratio of \fetfive/\fetsix\ $EW$s is at a maximum during dips
(orbital phase $\sim$0.8), consistent with the presence of
less-ionized material. \citet{1254:boirin03aa} did not examine the dip
seen from \twelve\ due to simultaneous enhanced background counting
rate related to solar activity. Despite the lower signal to noise
ratio implied during the dip, a re-analysis of the 2001 XMM-Newton
observation using the photo-ionized absorber model presented in this
paper (assuming the same ionizing continuum) indicates that \logxi\ is
$6.1 \pm 0.4 $, $5.1\,^{+0.4}_{-0.3} $ and $4.9\,^{+0.7}_{-0.4}$,
during persistent, shallow and deep dipping intervals, respectively,
clearly demonstrating that the absorber is less ionized during
dipping.  In the case of \src, this absorbing material has an
ionization parameter which decreases from a \logxi\ of $3.9 \pm 0.1$
during persistent intervals to $3.13 \pm 0.07$ during deep dipping.
We note that the absolute values of $\xi$ might suffer from a
systematic uncertainty due to mis-evaluation of the ionizing continuum
(see Appendix~A). However, this systematic effect does not affect the
relative differences in $\xi$ observed between persistent and dipping
intervals.  Thus, it is likely that the absorbing material responsible
for the dipping activity in X-ray binaries is in general less-ionized
than during persistent intervals.

In the case of \src, we demonstrate not only that the ionization
parameter is lower during dipping, but also that the equivalent
hydrogen column density is significantly higher than during persistent
intervals. The line-of-sight absorber may be considered to have
changed from being ``hot'' (almost fully ionized) to ``warm''. During
persistent emission, most of the abundant metals except for Fe are
almost fully ionized and the only prominent absorption features are
due to \fetfive\ and \fetsix\ (Fig.~\ref{fig:transmission}; upper
right panel). As the level of dipping increases, absorption from a
much wider range of ions is evident because of the higher column
density and the lower ionization parameter. The continuum transmission
becomes highly affected due to the continuum opacity of the ions and
to the presence of many more deep narrow features which blend together
(Fig.~\ref{fig:transmission}, middle and lower right panels). This
results in an apparent change in the continuum shape.

The spectral changes during dips from LMXBs are often modeled using
the ``progressive covering'', or ``complex continuum'' approach
\citep[e.g.,][]{1916:church97apj,1323:balucinska99aa,1323:barnard01aa}.
There, the X-ray emission is assumed to originate from a point-like
blackbody, or disk-blackbody component, together with a power-law
component from an extended corona.  This approach models the spectral
changes during dipping intervals by the partial and progressive
covering of the extended component by an opaque neutral absorber. In
the case of \src\ we have demonstrated that changes in the properties
of an ionized absorber provide an alternative explanation for the
overall spectral changes during dips.  The changes in the continuum
and in the narrow lines are modeled self-consistently in a simple way.
No partial covering of any component of the spectrum is required.
Consequently, the model does not exclude that the X-ray emission
originates from point-like components, and an extended corona is not
required.  Since similar changes in the Fe line ratios (and therefore
also in ionized absorbers) are observed from many other dip sources,
it is likely that this mechanism is important in modifying the
continuum shape during dips from these other sources as well.
However, answering whether this mechanism is sufficient to model the
complex changes seen from the other dip sources, or whether
``progressive covering'' is also required, will have to await further
investigation and particularly the high spectral resolution
observations of dipping LMXBs expected from Astro-E2.

While this paper was under revision, an analysis of the same
XMM-Newton observation of \src\ was reported by
\citet{1323:church05mnras}. There, it is argued that our model based
on an ionized absorber can be ruled out because ``the increase of
electron column density in dipping would cause a decrease in X-ray
continuum intensity at every energy due to Thomson scattering by a
factor $\exp[-N_{\rm e}\sigma_{\rm T}]$'', and in particular a too
large decrease of the 20--50~keV flux compared to what was observed
with BeppoSAX. However, the electron scattering optical depth,
$\tau_{\rm e}$ (see Sect.~\ref{sec:narrow}), is energy dependent, and
its approximation by the energy independent Thomson term, $N_{\rm
e}\sigma_{\rm T}$, does not hold at high energy. In fact, using the
numbers quoted in our Table~\ref{tab:weightedaverages}, we calculate
that our model (in which the energy dependent Klein-Nishina formula is
used to compute $\tau_{\rm e}$) predicts a 20--50~keV flux decrease of
$\sim$25\% during dips, consistent with the value of less than $10 \pm
10$\% measured with BeppoSAX \citep{1323:balucinska99aa}.

\begin{acknowledgements}
  We thank Frank Verbunt and Jean in~'t Zand for helpful discussions
  and careful reading of the manuscript.  We are grateful to the
  anonymous referee whose comments have lead to important improvements
  in the analysis.  This work is based on observations obtained with
  XMM-Newton, an ESA science mission with instruments and
  contributions directly funded by ESA member states and the USA
  (NASA).  The SRON is supported financially by the NWO, the
  Netherlands Organization for Scientific Research. \maria\
  acknowledges an ESA Fellowship.
\end{acknowledgements}

\appendix

\section{The photo-ionized absorber}

%
%
\begin{table}[!t]
\begin{center}
\caption[]{Column densities of the 20 first most abundant ions
  in the ionized absorber as predicted by \xabs\ in the persistent,
  shallow and deep dipping spectra fit with the {\tt
  abs*xabs*(pl+bb+gau)} model given in Table~\ref{tab:spectrum}. $R$=1
  corresponds to the highest column density. }
\begin{tabular}{ll@{\extracolsep{0.45cm}}r@{\extracolsep{0.25cm}}r@{\extracolsep{0.55cm}}r@{\extracolsep{0.25cm}}r@{\extracolsep{0.55cm}}r@{\extracolsep{0.25cm}}r}

\hline
\hline
\noalign {\smallskip}

\multicolumn{1}{l}{} & \multicolumn{3}{r}{Persistent} &  \multicolumn{4}{c}{Dipping}\\
\multicolumn{2}{c}{} & \multicolumn{2}{c}{} &  \multicolumn{2}{c}{Shallow}  & \multicolumn{2}{c}{Deep}\\
\noalign {\smallskip}
\multicolumn{2}{l}{Ion} & \logn & $R$ & \logn & $R$   &\logn & $R$  \\
\multicolumn{2}{c}{} & (cm$^{-2}$) &  &  (cm$^{-2}$) &   & (cm$^{-2}$) & \\
\noalign {\smallskip}
\hline
\noalign {\smallskip}

\multicolumn{2}{l}{\ion{H}{II}} & 22.55 & 1 & 22.74 & 1 & 23.14 & 1 \\
\multicolumn{2}{l}{\ion{He}{III}} & 21.48 & 2 & 21.67 & 2 & 22.07 & 2 \\
\multicolumn{2}{l}{\ion{C}{VII}} & 18.94 & 4 & 19.13 & 4 & 19.53 & 4 \\
\multicolumn{2}{l}{\ion{N}{VIII}} & 18.48 & 6 & 18.67 & 6 & 19.07 & 6 \\
\multicolumn{2}{l}{\ion{O}{VIII}} & -& -& 16.80 & 19 & 18.01 & 11 \\
\multicolumn{2}{l}{\ion{O}{IX}} & 19.24 & 3 & 19.43 & 3 & 19.82 & 3 \\
\multicolumn{2}{l}{\ion{Ne}{X}} & -& -& -& -& 17.92 & 15 \\
\multicolumn{2}{l}{\ion{Ne}{XI}} & 18.55 & 5 & 18.74 & 5 & 19.11 & 5 \\
\multicolumn{2}{l}{\ion{Na}{XII}} & 16.88 & 15 & 17.06 & 16 & -& - \\
\multicolumn{2}{l}{\ion{Mg}{XIII}} & 18.08 & 8 & 18.27 & 7 & 17.90 & 16 \\
\multicolumn{2}{l}{\ion{Al}{XIV}} & 17.01 & 13 & 17.19 & 14 & -& - \\
\multicolumn{2}{l}{\ion{Si}{XIV}} & 16.56 & 18 & 17.19 & 13 & 18.21 & 9 \\
\multicolumn{2}{l}{\ion{Si}{XV}} & 18.08 & 7 & 18.24 & 8 & 18.46 & 8 \\
\multicolumn{2}{l}{\ion{S}{XVI}} & 16.64 & 17 & 17.25 & 12 & 18.06 & 10 \\
\multicolumn{2}{l}{\ion{S}{XVII}} & 17.78 & 9 & 17.91 & 10 & 17.93 & 12 \\
\multicolumn{2}{l}{\ion{Ar}{XIX}} & 16.88 & 14 & 16.95 & 17 & -& - \\
\multicolumn{2}{l}{\ion{Ca}{XXI}} & 16.78 & 16 & -& -&  -& - \\
\multicolumn{2}{l}{\ion{Fe}{XXI}} & -& -& -& -& 17.62 & 20 \\
\multicolumn{2}{l}{\ion{Fe}{XXII}} & -& -& -& -& 17.88 & 17 \\
\multicolumn{2}{l}{\ion{Fe}{XXIII}} & -& -& -& -& 17.93 & 13 \\
\multicolumn{2}{l}{\ion{Fe}{XXIV}} & -& -& 17.11 & 15 & 17.80 & 18 \\
\multicolumn{2}{l}{\ion{Fe}{XXV}} & 17.43 & 11 & 17.92 & 9 & 17.92 & 14 \\
\multicolumn{2}{l}{\ion{Fe}{XXVI}} & 17.66 & 10 & 17.67 & 11 & -& - \\
\multicolumn{2}{l}{\ion{Fe}{XXVII}} & 17.41 & 12 & 16.94 & 18 & -& - \\
\multicolumn{2}{l}{\ion{Ni}{XXVII}} & 16.42 & 19 & -& -& -& - \\
\multicolumn{2}{l}{\ion{Ni}{XXVIII}} & 16.41 & 20 & -& -& -& - \\

\hline
\end{tabular} 
\label{tab:xabscolumns}
\end{center}
\end{table}

In this appendix, we give a detailed description of the photo-ionized
absorber as modeled by \xabs\ in the  averaged spectra of
\src\ (see Sects.~\ref{sec:spectralanalysis} and \ref{sec:averaged},
and Table~\ref{tab:spectrum}) and estimate the validity of some
assumptions of the \xabs\ model in the case of \src.

The electron effective temperature implied by the photo-ionization
model is $\sim$270~eV, $\sim$150~eV and $\sim$70~eV, for the
persistent, shallow and deep dipping spectra, respectively.
Table~\ref{tab:xabscolumns} lists the column densities of the most
abundant ions.  Table~\ref{tab:xabslines} lists the strongest
absorption lines due to the ionized absorber.
Table~\ref{tab:xabsedges} lists the strongest absorption edges. They
are significant only in the deep dipping spectrum. There, the \ews\ of
these edges are high compared with the strength of the lines from the
same ions.  This confirms that the ionized absorber is in a saturated
regime during deep dipping, as opposed to an unsaturated regime during
persistent and shallow dipping emission.

%
%
\begin{table}[!ht]
\begin{center}
\caption[]{
  \ews\ of the strongest absorption lines due to the ionized absorber
  as predicted by \xabs\ for the persistent, shallow and deep dipping
  spectra fit with the {\tt abs*xabs*(pl+bb+gau)} model given in
  Table~\ref{tab:spectrum}. For each spectrum, the predicted
  equivalent width, \ew, and the rank $R$ ($R$=1 for the line with the
  highest \ew) of each line is indicated.  The table is restricted to
  the 21 strongest lines for each of the three spectra. Therefore, the
  symbol~- does not mean that the line is not predicted, but that its
  \ew\ is smaller than the \ew\ of the 21$^{st}$ strongest line
  predicted for the given photo-ionized absorber. For each line, we
  indicate the quantum numbers of the electron involved in the
  transition, the upper term and the energy.}
\begin{tabular}{l@{\extracolsep{0.05cm}}l@{\extracolsep{0.2cm}}l@{\extracolsep{0.0cm}}r@{\extracolsep{0.3cm}}r@{\extracolsep{0.2cm}}r@{\extracolsep{0.3cm}}r@{\extracolsep{0.2cm}}r@{\extracolsep{0.3cm}}r@{\extracolsep{0.2cm}}r}

\hline
\hline
\noalign {\smallskip}
     &  &  &  & \multicolumn{2}{r}{Pers.} &  \multicolumn{4}{c}{Dipping}\\
\noalign {\smallskip}
& & & & &  &  \multicolumn{2}{c}{Sha.}  & \multicolumn{2}{c}{Deep}\\
\noalign {\smallskip}
\multicolumn{2}{l}{Ion}  & & Energy & \ew & $R$ &  \ew & $R$   & \ew & $R$ \\
\multicolumn{3}{r}{Transition}   & (keV) & (eV) &  &  (eV) &   & (eV) &  \\
\noalign {\smallskip}
\hline
\noalign {\smallskip}

\oeight\        & 1s-2p & \dptd & 0.654 &- &- & 1.6 & 19 &- &- \\
\neten\         & 1s-2p         & \dpud & 1.022         &- &- & 1.5 & 21 &- &- \\
\fetfour\       & 2s-3p & \dpud & 1.163 &- &- & 1.6 & 18 &- &- \\
\fetfour\       & 2s-3p & \dptd & 1.167         &- &- & 2.8 & 10 &- &- \\
\mgtwelve\      & 1s-2p         & \dptd & 1.473        & 0.4 & 20 & 1.7 & 17 &- &- \\
\sithirteen\      & 1s-2p              &  \upu   & 1.865 &- &- &- &-  & 3.0 & 21\\
\sifourteen\    & 1s-2p         & \dpud & 2.004 & 0.5 & 17 & 2.2 & 14 & 3.2 & 18 \\
\sifourteen\    & 1s-2p & \dptd & 2.006 & 1.1 & 10 & 4.0 & 9 & 3.6 & 16 \\
\sfifteen\      & 1s-2p & \upu  & 2.461        &- &- &- &- & 4.3 & 11\\
\ssixteen\      & 1s-2p & \dpud & 2.620 & 0.7 & 15 & 2.5 & 12 & 3.7 & 15\\
\ssixteen\      &1s-2p  & \dptd & 2.623 & 1.3 & 9 & 4.7 & 7 & 4.3 & 12 \\
\sfifteen\        &  1s-3p         & \upu  & 2.884 &- &- &- &- & 3.1 & 19 \\
\ssixteen\      & 1s-3p         & \dpud         & 3.107        &- &- &- &- & 3.1 & 20\\
\arseventeen\   & 1s-2p         & \upu  & 3.140        &- &- &- &- & 3.9 & 14\\
\canineteen\    & 1s-2p         & \upu  & 3.902        &- &- &- &- & 4.9 & 9\\
 \catwenty\     & 1s-2p & \dptd & 4.108 & 0.6 & 16 & 1.6 & 20 &- &- \\
\fetone\        & 1s-2p         & $^a$  & 6.544        &-      &-      &-      &-      & 8.3 & 6 \\
\fettwo\        & 1s-2p         &$^b$   & 6.586        &-      &-      &-      &-      & 8.7 & 5 \\
\fettwo\        & 1s-2p         &$^c$   & 6.587        &-      &-      &-      &-      & 9.1 & 4 \\
\fetthree\      & 1s-2p         &$^d$   & 6.596        &-      &-      &-      &-      &  3.5 & 17 \\
\fetthree\      & 1s-2p &$^e$   & 6.629        &-      &-      & 1.9 & 16 & 11.7 & 2 \\
\fetfour\       & 1s-2p &$^f$   & 6.653        &-      &-      & 2.2 & 15 & 6.0 & 8 \\
\fetfour\       & 1s-2p &$^g$   & 6.662 & 0.8 & 14 & 6.4 & 5 & 9.5 & 3 \\
\fetfour\       & 1s-2p &$^h$   & 6.677        &- &- &- &- & 4.5 & 10 \\
\fetfive\       & 1s-2p         & \upu  & 6.700 & 21.9 & 1 & 38.2 & 1 & 11.9 & 1 \\
\fetsix\        & 1s-2p         & \dpud & 6.952        & 6.8 & 3 & 6.6 & 4 &- &- \\
\fetsix\        & 1s-2p         & \dptd & 6.973        & 13.3 & 2 & 12.3 & 3 &- &- \\
\nitseven\      & 1s-2p         & \upu  & 7.805        & 1.9 & 6 & 4.2 & 8 & - &- \\
\fetfive\       & 1s-3p         & \upu  & 7.881        & 4.6 & 4 & 12.5 & 2  & 7.3 & 7 \\
\niteight\      & 1s-2p  & \dpud         & 8.073        & 0.4 & 21 &- &- &- &- \\
\niteight\      & 1s-2p         & \dptd & 8.101        & 0.8 & 13 &- &- &- &- \\
 \fetsix\       & 1s-3p         & \dpud & 8.246        & 1.3 & 8 &- &- &- &-\\
 \fetsix\       & 1s-3p         & \dptd & 8.252        & 2.6 & 5 & 2.7 & 11 &- &- \\
 \fetfive\      & 1s-4p         & \upu  & 8.296        & 1.7 & 7 & 5.1 & 6 & 4.0 & 13 \\
\fetfive\       & 1s-5p & \upu  & 8.487        &  0.8 & 12 & 2.5 & 13 &- &- \\
 \fetsix\       & 1s-4p         & \dpud & 8.698        & 0.5 & 18 &- &- &- &- \\
\fetsix\        & 1s-4p & \dptd & 8.701       & 1.0 & 11 &- &- &- &-\\
\fetsix\         & 1s-5p          & \dptd          & 8.909  & 0.5 & 19 &- &- &- &- \\
\noalign {\smallskip}
\hline
\noalign {\smallskip}
\multicolumn{10}{l}{$^a$ $\rm 1s\,2s^2\,2p^3$ \tpu, $^b$ $\rm 1s\,2s^2\,2p^2$ \dpud, $^c$ $\rm 1s\,2s^2\,2p^2$ \tpu, }\\
\multicolumn{10}{l}{$^d$ $\rm 1s\,2s^2\,2p$ \tpu, $^e$ $\rm 1s\,2s^2\,2p$ \upu, $^f$ $\rm 1s\,2s\,2p\, (^3P)$ \dpud,}\\
\multicolumn{10}{l}{$^g$ $\rm 1s\,2s\,2p\, (^3P)$ \dptd, $^h$ $\rm 1s\,2s\,2p\, (^1P)$ \dpud.}\\
\end{tabular}

\label{tab:xabslines}
\end{center}
\end{table}

To evaluate how the choice of the ionizing continuum affects our
results, we calculated new grids of relative ionic abundances using
CLOUDY assuming various ionizing spectra, and fit the average spectra
again using these new grids.  When the ionizing spectrum is assumed to
be a cutoff power-law with \phind\ of 2.1 (instead of 1.96) and \ecut\
of 44~keV, we find that the best-fit parameters are all consistent
with those in Table~\ref{tab:spectrum} except $\xi$ which is higher,
with \logxi~$= 4.13 \pm 0.09$ in the case of the persistent emission.
Indeed, the considered ionizing spectrum is softer, and the electron
temperature of the photo-ionized absorber has dropped to
$\sim$230~eV. Therefore a higher luminosity, and hence a higher $\xi$
(if $n_{\rm e}$ and $r^{2}$ are unchanged), are required to produce
the iron ions in the same ratios.  On the contrary, when the assumed
ionizing spectrum consists of a cutoff power-law with \phind\ of 1.7
and \ecut\ of 44~keV, $\xi$ becomes significantly lower, with
\logxi~$= 3.63 \pm 0.07$ for the persistent spectrum, as expected from
a harder ionizing spectrum (the electron temperature is now
$\sim$330~eV).  For the shallow and deep dipping spectrum, we derive
\logxi\ of $3.4 \pm 0.1$, $2.95 \pm 0.15$, respectively.  Therefore,
although the absolute value of $\xi$ depends on the assumed shape of
the ionizing spectrum (which is poorly known), the relative change of
$\xi$ observed from persistent to deep dipping is un-affected.  Note
that the absolute value of $\xi$ does not only depend on the assumed
ionizing spectrum, but more generally, on the photo-ionization model
(XSTAR or CLOUDY) used (Steenbrugge et al. 2005).
\nocite{5548:steenbrugge05aa}

Using the \xabs\ model to account for narrow absorption lines, we have
neglected resonant emission lines coming from ionized material located
outside the line-of-sight, that would partly re-fill-in the absorption
features. If these emission lines were not neglected, the observed
equivalent width of a feature, $EW_{\rm o}$, would be related to the
equivalent width, $EW_{\rm a}$, due to absorption only, by the
relation: $EW_{\rm o} = EW_{\rm a} (1 - \Omega/4\pi)$, where $\Omega$
is the solid angle subtended by the ionized material. The column
density of the absorber should then be estimated from $EW_{\rm a}$
rather than from $EW_{\rm o}$.  Assuming that the ionized material
subtends a solid angle $\Omega$ of $0.2 \times 4\pi$ \citep[see
e.g. ][]{gx13:ueda04apj}, we find that for the persistent emission
(linear regime of the curve-of-growth), the equivalent column density
of the absorber would be higher by 25\%. This value is comparable to
the uncertainty on \nhxabs\ in the average persistent spectrum
(Table~\ref{tab:spectrum}), and the standard deviation of the \nhxabs\
values among the individual persistent segments
(Fig.~\ref{fig:parameters}).

%
%
\begin{table}[!h]
\begin{center}
\caption[]{
Optical depths, $\tau$, and $EW$s of the strongest absorption edges as
predicted by \xabs\ in the persistent, shallow and deep dipping
spectra (see Table~\ref{tab:spectrum}).  $R$=1 corresponds to the edge
with the highest \ew.}
\begin{tabular}{ll@{\extracolsep{-0.25cm}}r@{\extracolsep{0.25cm}}r@{\extracolsep{0.1cm}}r@{\extracolsep{0.1cm}}r@{\extracolsep{0.45cm}}r@{\extracolsep{0.1cm}}r@{\extracolsep{0.1cm}}r@{\extracolsep{0.45cm}}r@{\extracolsep{0.1cm}}r@{\extracolsep{0.25cm}}r}

\hline
\hline
\noalign {\smallskip}
\multicolumn{2}{c}{} & \multicolumn{4}{r}{Persistent} &  \multicolumn{6}{c}{Dipping}\\
\multicolumn{3}{c}{} & \multicolumn{3}{c}{} &  \multicolumn{3}{c}{Shallow}  & \multicolumn{3}{c}{Deep}\\
\noalign {\smallskip}
\multicolumn{3}{l}{Edge\hspace{0.1cm} Energy} & \multicolumn{3}{c}{$\tau$ \hfill \ew\ \hfill $R$} & \multicolumn{3}{c}{$\tau$ \hfill \ew\ \hfill $R$} & \multicolumn{3}{c}{$\tau$ \hfill \ew\ \hfill $R$}\\
\multicolumn{2}{c}{} & (keV) & \multicolumn{3}{c}{(eV)}  & \multicolumn{3}{c}{(eV)}   &\multicolumn{3}{c}{(eV)}  \\
\noalign {\smallskip}
\hline
\noalign {\smallskip}

\multicolumn{2}{l}{\ion{Si}{ xiv}}   & 2.67        &-        &-   &-      &-       &-   &-    & 0.052         & 76  & 4 \\
\multicolumn{2}{l}{\ion{Fe}{ xxii}}  & 8.38        &-      &-      &-      &-      &-      &-      & 0.017         & 82  & 2 \\
\multicolumn{2}{l}{\ion{Fe}{ xxiii}}  & 8.48        &-      &-      &-      &-      &-      &-      & 0.018         & 90  & 1 \\
\multicolumn{2}{l}{\ion{Fe}{ xxv}}   & 8.83        & 0.005        & 26  & 1     & 0.016        & 81  & 1     & 0.016         &   81       & 3     \\
\multicolumn{2}{l}{\ion{Fe}{ xxvi}}  & 9.28        & 0.004       & 22  & 2     & 0.004        & 22  & 2     &-      &-      &-      \\
  
\hline

\end{tabular}
\label{tab:xabsedges}
\end{center}
\end{table}

Using \xabs, we have also neglected recombination lines from the
 photo-ionized material.  If such line emission from the ions were
 important, not only the resonant line would be emitted, but also the
 associated intercombination and forbidden lines, in the case of
 He-like ions.  In a photo-ionized plasma, the \fetfive\ forbidden
 emission line at 6.6341~keV (or alternatively, the intercombination
 line if the density is high) is expected to be a factor $\sim$2.3
 stronger than the resonance emission line at 6.6986~keV. Now, no
 narrow line is detected at the energy of the forbidden transition in
 the average persistent spectrum of \src. This indicates that narrow
 line emission does not play an important role compared to narrow line
 absorption. To further quantify this, we have derived the upper-limit
 on the normalization of the forbidden line, $k_z$, and added a narrow
 {\it emission} line with an energy fixed to that of the resonant
 transition and with a normalization fixed to $k_z$/2.3, to the model
 described in Table~\ref{tab:spectrum} for the persistent average
 spectrum. We then performed the fit again and found that all the
 best-fit parameters are consistent with the previous values.  In
 particular, \nhxabs\ is found to be $3.6\,^{+1.0}_{-0.8}$~\ttnh, in
 agreement with the value reported in Table~\ref{tab:spectrum}.
 Therefore, the assumption that recombination lines from the
 photo-ionized material can be neglected is \emph{a posteriori}
 verified in the case of \src.


\bibliographystyle{aa}

\end{document}